\documentclass[preprint,english,aps,prb,superscriptaddress,showpacs,floatfix]{revtex4}
\usepackage{graphicx}
\usepackage{amsmath,amssymb,amsfonts}
\usepackage{natbib}

\makeatletter

\usepackage{babel}
\makeatother
\begin{document}

\title{Mapping quantum-classical Liouville equation: projectors and trajectories}

\author{Aaron Kelly}
\email{atkelly@stanford.edu}
\affiliation{Chemical Physics Theory Group, Department of Chemistry, University
of Toronto, Toronto, ON M5S 3H6, Canada}
\affiliation{Department of Chemistry, Stanford University, 333 Campus Drive, Stanford, CA 94305, USA}

\author{Ramses van Zon}
\email{rzon@scinet.utoronto.ca}
\affiliation{Chemical Physics Theory Group, Department of Chemistry, University
of Toronto, Toronto, ON M5S 3H6, Canada}
\author{Jeremy Schofield}
\email{jmschofi@chem.utoronto.ca}
\affiliation{Chemical Physics Theory Group, Department of Chemistry, University
of Toronto, Toronto, ON M5S 3H6, Canada}
\author{Raymond Kapral}
\email{rkapral@chem.utoronto.ca}
\affiliation{Chemical Physics Theory Group, Department of Chemistry, University
of Toronto, Toronto, ON M5S 3H6, Canada}

\date{\today{}}

\begin{abstract}

The evolution of a mixed quantum-classical system is expressed in the mapping formalism where discrete quantum states are mapped onto oscillator states,
resulting in a phase space description of the quantum degrees of freedom. By defining projection operators onto the
mapping states corresponding to the physical quantum states, it is shown that the mapping quantum-classical Liouville operator commutes
with the projection operator so that the dynamics is confined to the physical space. It is also shown that a trajectory-based
solution of this equation can be constructed that requires the simulation of an ensemble of entangled trajectories. An approximation
to this evolution equation which retains only the Poisson bracket contribution to the evolution operator does admit a solution
in an ensemble of independent trajectories but it is shown that this operator does not commute with the projection operators
and the dynamics may take the system outside the physical space. The dynamical instabilities, utility and domain of validity
of this approximate dynamics are discussed. The effects are illustrated by simulations on several quantum systems.

\end{abstract}

\maketitle

\section{Introduction} \label{sec:intro}

Since phenomena such as electron and proton transfer dynamics~\cite{bell73,kornyshev97}, excited state relaxation processes~\cite{femtochem01} and energy transport in light harvesting systems~\cite{fleming09,scholes10} are quantum in nature, the development of theoretical descriptions and simulation methods for quantum many-body systems is a central topic of research. Although various techniques can be used to study such problems, quantum-classical methods~\cite{herman94,0chap-tully98,0book-billing03,truhlar04,kapral06,subotnik11}, where certain degrees of freedom are singled out for a full quantum treatment while other environmental variables are treated classically, permit one to investigate large and complex systems that cannot be studied by other means.

In this article we consider descriptions of the dynamics based on the quantum-classical Liouville
equation~\cite{kapral06} (QCLE) and, in particular, its representation in the mapping
basis~\cite{kim-map08,nassimi09,nassimi10}. The mapping formalism provides an exact mapping of
discrete quantum states onto continuous variables~\cite{stock05} and in quantum-classical systems leads to phase-space-like evolution equations for both quantum and classical degrees of freedom. The mapping basis has been used in a number of different quantum-classical formulations, often based on semi-classical path integral expressions for the dynamics~\cite{miller78,meyer79,sun98,miller01,miller07,stock97,muller98,thoss99,thoss04,stock05,bonella01,bonella03,bonella05,dunkel08}. The representation of the quantum-classical Liouville equation in the mapping basis leads to an equation of motion whose Liouvillian consists of a Poisson bracket term in the full quantum subsystem-classical bath phase space, and a more complex term involving second derivatives of quantum phase space variables and first derivatives with respect to bath momenta~\cite{kim-map08}. This latter contribution has been shown to be an excess coupling term related to a portion of the back reaction of the quantum subsystem on the bath~\cite{nassimi10}.

Various aspects of the QCLE in the mapping basis and properties of its full and approximate solutions are discussed in this paper.
The solutions of the quantum-classical Liouville equation cannot be obtained from the dynamics of an ensemble of independent classical-like trajectories~\cite{kapral99}. In the adiabatic basis this equation admits a solution in terms of surface-hopping trajectories~\cite{kapral99,mackernan08,sergi03b}, but other schemes have been used to simulate the dynamics~\cite{martens96,donoso98,wan00,wan02}. When it is expressed in the mapping basis, we show that a solution can be obtained in terms of an ensemble of entangled trajectories. The excess coupling gives rise to correlations between the dynamics of the quantum mapping degrees of freedom and the bath phase space variables that are responsible for the entanglement of the trajectories in the ensemble. The derivation of the entangled trajectory picture is similar to that for trajectory solutions of the Wigner-Liouville equation~\cite{donoso01,donoso03}.

If the excess coupling term is dropped and only the Poisson bracket part of the Liouvillian is retained, a very simple equation of motion that admits a solution in terms of characteristics is obtained. Consequently, its solutions can be obtained from simulations of an ensemble of independent trajectories evolving under Newtonian dynamics. The set of ordinary differential equations has appeared earlier in mapping formulations based on semi-classical path integral formulations of the dynamics~\cite{stock05,miller01,miller07}, indicating a close connection between this approximation to the quantum-classical Liouville equation and those formulations. The solutions of this Poisson bracket approximation to the QCLE, as well as those of other semi-classical schemes that use this set of evolution equations, often provide a quantitatively accurate description of the dynamics~\cite{stock05,kim-map08,nassimi10}.   However for some systems the solutions are not without artifacts and difficulties. Some of these difficulties can be traced to the fact that the independent-ensemble dynamics can take the system out of the physical space and inverted potentials can appear in the evolution equations, which may lead to instabilities~\cite{stock05,bonella01,bonella05}.

The main results of this paper are as follows: We present derivations of expressions for mapping quantum-classical (MQCL) evolution equations
and expectation values of operators that explicitly show how projection operators onto the physical mapping eigenstates enter the formulation. We demonstrate that the MQCL operator commutes with this projection operator so that dynamics under this evolution is confined to the physical space. This full quantum-classical dynamics in the mapping basis can be simulated by an ensemble of entangled trajectories. We also show that when the excess coupling term is neglected the resulting Poisson bracket operator no longer commutes with the projection operator so that this approximate dynamics can take the system out of the physical space. Given this context, we revisit the issue of instabilities in the dynamics of the Poisson bracket approximation and discuss the conditions under which such instabilities are likely to arise and lead to inaccuracies in the solutions.

In Sec.~\ref{sec:qcl-map} we outline the representation of the
quantum-classical Liouville equation in the mapping basis and show how average values of time dependent
observables may be computed. We also define a projection operator onto the mapping states and show how this projector enters
the expressions for the expectation values and evolution equations.  Section~\ref{sec:traj} briefly describes the
entangled trajectory solution to the QCLE in the mapping basis. This section also shows that when the excess coupling
term is neglected, a solution in terms of an ensemble of independent
trajectories is possible.  In Sec.~\ref{sec:PBME} the approximate evolution equation obtained by retaining only
the Poisson bracket term in the Liouville operator is considered and the dynamical instabilities that can arise in the course of the
evolution are highlighted. Various aspects of the theoretical analysis that concern the approximate solutions and resulting instabilities are illustrated by simulations of a number of model systems. A brief summary of the main results of the study, along with comments, are given in
Sec.~\ref{sec:conc}. The Appendices provide material to support the text. In particular, we describe an efficient simulation algorithm
for the ordinary differential equations that underlie the solutions of the Poisson bracket approximation to the QCLE.

\section{Quantum-Classical Liouville Equation: Mapping, Projectors and Expectation Values}\label{sec:qcl-map}

The quantum-classical Liouville equation (QCLE),
\begin{eqnarray} \label{eq:qcl-rho}
  && \frac{\partial }{\partial t}\hat{\rho}_W (X, t) =-i  \hat{{\mathcal L}}\hat{\rho}_W (X,t),
\end{eqnarray}
describes the time evolution of the density matrix $\hat{\rho}_W(X,t)$, which is a quantum operator that depends on the classical phase space variables $X=(R,P)=(R_1,R_2,...,R_{N_e},P_1,P_2,...,P_{N_e})$ of the environment. The quantum-classical Liouville operator is defined by
\begin{equation}\label{eq:qclop-abs}
i\hat{{\mathcal L}} \cdot = \frac{i}{\hbar}[\hat{H}_W,\cdot] - \frac{1}{2}(\{\hat{H}_W,\cdot\}
-\{\cdot,\hat{H}_W\}),
\end{equation}
where $\hat{H}_W(X)$ is the partial Wigner transform of the total Hamiltonian of the system, $[\cdot,\cdot]$ is the commutator and $\{\cdot,\cdot\}$ is the Poisson bracket in the phase space of the classical variables $X$. The total Hamiltonian may be written as the sum of environmental (bath), subsystem and coupling terms, $\hat{H}_W(X)=
H_e(X)+\hat{h}_s+\hat{V}_c(R)$, where $H_e(X)=P^2/2M+V_e(R)$ is the bath Hamiltonian
with $V_e(R)$ the bath potential energy, $\hat{h}_s=\hat{p}^2/2m+\hat{V}_s$ is the
subsystem Hamiltonian with $\hat{p}$ and $\hat{V}_s$ the subsystem momentum and
potential energy operators, and $\hat{V}_c(R)$ is the coupling potential
energy operator. Here $m$ and $M$ are the masses of the subsystem and bath particles,
respectively.

The QCLE may be written in the basis, $\{|\lambda\rangle ; \lambda =1, \dots, N\}$, that spans the quantum subsystem space with
eigenfunctions defined by the eigenvalue problem,
$\hat{h}_s |\lambda\rangle =\epsilon_\lambda |\lambda\rangle$. Taking matrix elements of Eq.~(\ref{eq:qcl-rho}) we obtain
\begin{eqnarray} \label{eq:qcl-rho-sub}
  && \frac{\partial }{\partial t}{\rho}^{\lambda \lambda'}_W (X, t) =-i  {{\mathcal L}}_{\lambda \lambda', \nu \nu'}{\rho}^{\nu \nu'}_W (X,t).
\end{eqnarray}
The Einstein summation convention is used here and in subsequent equations although, on occasion, sums will be explicitly written for
purposes of clarity. The QCL operator in the subsystem basis is~\cite{kapral99}
\begin{eqnarray}\label{eq:QCLop-sub}
&&i{\cal L}_{\lambda \lambda',\nu \nu'}=
i \omega_{\lambda \lambda'} \delta_{\lambda \nu}
\delta_{\lambda' \nu'}
-\frac{i}{\hbar}
\left(\delta_{\lambda \nu}  V_c^{\nu' \lambda'}
- V_c^{\lambda \nu} \delta_{\lambda' \nu'} \right) \nonumber \\
&&\qquad + \left(\frac{P}{M}\cdot{\partial \over \partial R} +F_e(R)\cdot
{\partial \over \partial P} \right) \delta_{\lambda \nu}
\delta_{\lambda' \nu'} \nonumber \\
&& \qquad -
\frac{1}{2} \left(\delta_{\lambda' \nu'} {\partial V_c^{\lambda \nu}
\over \partial R} + \delta_{\lambda \nu}
{\partial V_c^{\nu' \lambda'} \over \partial R}
\right)\cdot{\partial  \over \partial P} \;,
\label{superLsub}
\end{eqnarray}
where $\omega_{\lambda \lambda'}=(\epsilon_\lambda-\epsilon_{\lambda'})/\hbar$ and $F_e(R)=-\partial V_e /\partial R$ is the force due to
molecules in the environment.

The evolution equation for an observable $\hat{B}_W(X)$, analogous to Eq.~(\ref{eq:qcl-rho}), is
\begin{eqnarray} \label{eq:qcl-obs}
  && \frac{d }{d t}\hat{B}_W (X, t) =i  \hat{{\mathcal L}}\hat{B}_W (X,t),
\end{eqnarray}
and its representation in the subsystem basis is analogous to Eq.~(\ref{eq:qcl-rho-sub}) with a change in sign on the right side.

\subsection{Representation in Mapping Basis and Projection Operators}

In the mapping basis~\cite{chap-schwinger65,stock05} the $|\lambda\rangle$ eigenfunctions of an $N$-state quantum subsystem
can be replaced with eigenfunctions of $N$ fictitious harmonic
oscillators, $|m_{\lambda}\rangle$, having occupation numbers which are limited
to 0 or 1: $|\lambda\rangle\rightarrow|m_{\lambda}\rangle=|0_{1},
\cdots,1_{\lambda},\cdots0_{N}\rangle$. Creation and annihilation operators on these states, $\hat{a}_{\lambda}^{\dag}$ and $\hat{a}_{\lambda}$, respectively, may be defined. For any operator $\hat{B}_W(X)$ whose matrix elements in the subsystem basis are $B^{\lambda \lambda'}_W(X)$,
we may associate a mapping basis operator $\hat{B}_W(X) \to \hat{B}_{m}(X)$, where
\begin{equation}
\hat{B}_{m}(X)=B_{W}^{\lambda\lambda'}(X)
\hat{a}_{\lambda}^{\dag}\hat{a}_{\lambda'}.
\label{Bm_sub}
\end{equation}
It is then evident that the matrix element
$B_{W}^{\lambda\lambda'}(X) =\langle\lambda|\hat{B}_{W}(X)|\lambda'\rangle=\langle
m_{\lambda}|\hat{B}_{m}(X)|m_{\lambda'}\rangle$.

The expression for $B_W^{\lambda \lambda'}(X)$ may also be written in terms of the Wigner transforms in the space of the mapping variables. Inserting complete sets of
coordinate states $\{|q\rangle, |q'\rangle \}$, and making the usual coordinate transformations appropriate for Wigner transforms, $(q,q') \to (r-z/2,r+z/2)$,
we obtain
\begin{eqnarray} \label{eq:expect-map}
&&B_{W}^{\lambda\lambda'}(X) =\langle m_{\lambda}|\hat{B}_{m}(X)|m_{\lambda'}\rangle =\\
&&\int dr dz\; \langle m_{\lambda}|r-\frac{z}{2} \rangle \langle r-\frac{z}{2}  |\hat{B}_{m}(X)|r+\frac{z}{2} \rangle \langle r+\frac{z}{2}|m_{\lambda'}\rangle \nonumber
\end{eqnarray}
Another form for the matrix element can be obtained by inserting the Wigner transform of an operator and its inverse as
\begin{eqnarray}\label{eq:wigner-op}
&&\langle r-\frac{z}{2}  |\hat{B}_{m}(X)|r+\frac{z}{2} \rangle = \frac{1}{(2\pi \hbar)^N} \int dp\; e^{-ip\cdot z/\hbar}
B_m({\mathcal X}),\nonumber \\
&&B_{m}({\mathcal X})= \int dz\; e^{ip\cdot z/\hbar}
\langle r-\frac{z}{2}|\hat{B}_{m}(X)|r+\frac{z}{2}\rangle.
\end{eqnarray}
Here ${\mathcal X}=(x,X)$ are the extended phase space coordinates for the subsystem mapping variables, $x=(r,p)=(r_1,...,r_N,p_1,...,p_N)$, and the environment, $X=(R,P)$. Making these substitutions in Eq.~(\ref{eq:expect-map}) we obtain,
\begin{equation} \label{eq:Bsubing}
B_{W}^{\lambda\lambda'}(X)= \int dx \; B_m({\mathcal X}) g_{\lambda \lambda'}(x),
\end{equation}
where we have defined~\cite{footnote:notation-change}
\begin{eqnarray}\label{eq:g-def}
&&g_{\lambda \lambda'}(x)=\frac{1}{(2\pi \hbar)^N}\int dz\; e^{-ip\cdot z/\hbar}
\langle r+\frac{z}{2}|m_{\lambda'}\rangle \langle m_{\lambda}|r-\frac{z}{2} \rangle \nonumber\\
&&\quad=\frac{1}{(2\pi \hbar)^N}\int dz\; e^{ip\cdot z/\hbar}
\langle r-\frac{z}{2}|m_{\lambda'}\rangle \langle m_{\lambda}|r+\frac{z}{2} \rangle.
\end{eqnarray}
Evaluating the integral we obtain an explicit expression for $g_{\lambda \lambda'}(x)$:
\begin{eqnarray} \label{eq:g-expression}
&&g_{\lambda \lambda'}(x)=\phi(x)\\
&&
\times \frac{2}{\hbar}
\Big[r_\lambda r_{\lambda'}+ p_{\lambda}p_{\lambda'} - i(r_\lambda p_{\lambda'} - r_{\lambda'} p_{\lambda})  - \frac{\hbar}{2}\delta_{\lambda\lambda'} \Big] , \nonumber
\end{eqnarray}
where $\phi(x)=(\pi \hbar)^{-N}\exp{(-x^2/\hbar)}$ is a normalized Gaussian function.  Here $x^2 =  r_\lambda r_\lambda
+ p_\lambda p_\lambda$ in the Einstein summation convention.

The expression for $B_m({\mathcal X})$ in Eq.~(\ref{eq:wigner-op})
can be simplified by evaluating the integral in the Wigner transform.
Using the definition of $\hat{B}_{m}(X)$ in Eq.~(\ref{Bm_sub}), Eq.~(\ref{eq:wigner-op}) may be written as
\begin{equation}
B_{m}({\mathcal X})=B_{W}^{\lambda\lambda'}(X) \int dz\; e^{ip\cdot z/\hbar}
\langle r-\frac{z}{2}|\hat{a}_{\lambda}^{\dag}\hat{a}_{\lambda'}|r+\frac{z}{2}\rangle.
\end{equation}
Noting that the factor multiplying $B_{W}^{\lambda\lambda'}(X)$ is the Wigner transform of $\hat{a}^\dagger_\lambda  \hat{a}_{\lambda'}$,
$(\hat{a}^\dagger_\lambda  \hat{a}_{\lambda'})_W(x)\equiv c_{\lambda \lambda'}(x)$, whose explicit value is
\begin{eqnarray}\label{eq:c-expression}
c_{\lambda \lambda'}(x)=\frac{1}{2\hbar}
[r_\lambda r_{\lambda'}+ p_{\lambda}p_{\lambda'} + i(r_\lambda p_{\lambda'} - r_{\lambda'} p_{\lambda})
- \hbar\delta_{\lambda\lambda'}],\nonumber \\
\end{eqnarray}
we find
\begin{eqnarray}\label{Bm_sub_wigner}
B_{m}({\mathcal X})=B_{W}^{\lambda\lambda'}(X)c_{\lambda \lambda'}(x).
\end{eqnarray}

We may deduce a number of other relations given the definitions stated above. A mapping operator $\hat{B}_{m}(X)$ acts on
mapping functions $|m_\lambda \rangle$. In this space we have the completeness relations $\hat{{\mathcal P}}=\sum_{\lambda=1}^N |m_\lambda \rangle \langle m_\lambda|=1$, where $\hat{{\mathcal P}}$ is projector onto the complete set of mapping states.~\cite{footnote:proj-equib} Thus, a mapping operator can be written using this projector as
\begin{eqnarray} \label{B-proj}
\hat{B}_{m}^{{\mathcal P}}(X)=\hat{{\mathcal P}} \hat{B}_{m}(X) \hat{{\mathcal P}} &=&|m_\lambda\rangle \langle
m_{\lambda}|\hat{B}_{m}(X)|m_{\lambda'}\rangle \langle m_{\lambda'}|\nonumber \\
&=&|m_\lambda\rangle {B}_W^{\lambda \lambda'}(X) \langle m_{\lambda'}|,
\end{eqnarray}
where in the second line we used the equivalence between matrix elements in the subsystem and mapping representations given in
Eq.~(\ref{eq:expect-map}). We can make use of the Wigner transforms defined in Eq.~(\ref{eq:wigner-op}) to write these relations in other forms.
Using the first equality in Eq.~(\ref{B-proj}) we have
\begin{eqnarray}\label{eq:proj-B}
&&{B}_{m}^{{\mathcal P}}({\mathcal X})= \int dz\; e^{ip\cdot z/\hbar}
\langle r-\frac{z}{2}|\hat{B}^{{\mathcal P}}_{m}(X)|r+\frac{z}{2}\rangle = \nonumber\\
&&\int dz\; e^{ip\cdot z/\hbar}
\langle r-\frac{z}{2}|m_\lambda\rangle \langle
m_{\lambda}|\hat{B}_{m}(X)|m_{\lambda'}\rangle \langle m_{\lambda'}|r+\frac{z}{2}\rangle \nonumber \\
&& = (2 \pi \hbar)^N g_{\lambda' \lambda}(x) \int dx' \; g_{\lambda \lambda'}(x') B_m(x',X), \nonumber \\
&& \equiv {\mathcal P} B_m({\mathcal X}),
\end{eqnarray}
where we used Eqs.~(\ref{eq:expect-map}) and (\ref{eq:Bsubing}). The last line defines the projection operator ${\mathcal P}$ that projects any function of the
mapping phase space coordinates, $f(x)$, onto the mapping states,
\begin{equation}\label{eq:proj-op}
{\mathcal P} f(x) = (2 \pi \hbar)^N g_{\lambda' \lambda}(x) \int dx' \; g_{\lambda \lambda'}(x') f(x').
\end{equation}
One may verify that ${\mathcal P}^2={\mathcal P}$ since
\begin{equation}\label{eq:gg-int}
(2 \pi \hbar)^N \int dx \; g_{\lambda \lambda'}(x) g_{\nu' \nu}(x) =\delta_{\lambda \nu} \delta_{\lambda' \nu'}.
\end{equation}

An equivalent expression for ${B}_{m}^{{\mathcal P}}({\mathcal X})$ can be obtained by starting with the last equality in Eq.~(\ref{B-proj})
and taking Wigner transforms to find
\begin{eqnarray} \label{Bm_sub_wigner-proj}
&&{B}_{m}^{{\mathcal P}}({\mathcal X})=\int dz\; e^{ip\cdot z/\hbar}
\langle r-\frac{z}{2}|m_\lambda\rangle {B}_W^{\lambda \lambda'}(X) \langle m_{\lambda'}|r+\frac{z}{2}\rangle \nonumber \\
&& \qquad \quad = (2 \pi \hbar)^N g_{\lambda' \lambda}(x) {B}_W^{\lambda \lambda'}(X).
\end{eqnarray}
This result also follows from Eq.~(\ref{eq:proj-B}) by substituting Eq.~(\ref{Bm_sub_wigner}) for $B_m({\mathcal X})$ and using the
fact that
\begin{equation} \label{eq:gc-int}
\int dx \; g_{\lambda \lambda'}(x) c_{\nu \nu'}(x) =\delta_{\lambda \nu} \delta_{\lambda' \nu'}.
\end{equation}
Finally, in view of the definition of the projection operator ${\mathcal P}$, in place of Eq.~(\ref{eq:Bsubing}) we may write
\begin{equation} \label{eq:Bsubing-proj}
B_{W}^{\lambda\lambda'}(X)= \int dx \; B_m^{{\mathcal P}}({\mathcal X}) g_{\lambda \lambda'}(x).
\end{equation}

An analogous set of relations apply to the matrix elements of the density operator, $\rho^{\lambda \lambda'}_W(X) =\langle\lambda|\hat{\rho}_{W}(X)|\lambda'\rangle=\langle m_{\lambda}|\hat{\rho}_{m}(X)|m_{\lambda'}\rangle$, where $\hat{\rho}_{m}(X)=\rho_{W}^{\lambda\lambda'}(X) \hat{a}_{\lambda}^{\dag}\hat{a}_{\lambda'}$. Taking the Wigner transform of
$\hat{\rho}_{m}(X)$ we find
\begin{eqnarray}\label{eq:unproj-den}
\rho_m({\mathcal X}) &=& \frac{1}{(2\pi \hbar)^N} \int dz\; e^{ip\cdot z/\hbar}
\langle r-\frac{z}{2}|\hat{\rho}_{m}(X)|r+\frac{z}{2}\rangle \nonumber \\
&&= \frac{1}{(2\pi \hbar)^N} {\rho}_W^{\lambda \lambda'}(X) c_{\lambda \lambda'}(x).
\end{eqnarray}
Likewise, starting from the expression for the projected density,
\begin{eqnarray}
\hat{\rho}^{{\mathcal P}}_m(X)&=&|m_\lambda\rangle \langle
m_{\lambda}|\hat{\rho}_{m}(X)|m_{\lambda'}\rangle \langle m_{\lambda'}|\nonumber \\
&=&|m_\lambda\rangle {\rho}_W^{\lambda \lambda'}(X) \langle m_{\lambda'}|,
\end{eqnarray}
its Wigner transform is
\begin{eqnarray}\label{eq:inversewigner-rho}
\rho_{m}^{{\mathcal P}}({\mathcal X})&=& \frac{1}{(2\pi \hbar)^N}\int dp\; e^{ip\cdot z/\hbar}
\langle r-\frac{z}{2}|\hat{\rho}_{m}^{{\mathcal P}}(X)|r+\frac{z}{2}\rangle \nonumber \\
&&= {\mathcal P} \rho_m({\mathcal X}),
\end{eqnarray}
which, repeating the steps that gave Eq.~(\ref{Bm_sub_wigner-proj}), yields
\begin{equation}\label{eq:proj-den}
{\rho}^{{\mathcal P}}_m({\mathcal X})= {\rho}_W^{\lambda \lambda'}(X) {g}_{\lambda' \lambda}(x).
\end{equation}
Following the analysis given above that led to Eq.~(\ref{eq:Bsubing}) for an operator, and using  the relation
\begin{equation}
\langle r-\frac{z}{2}  |\hat{\rho}_{m}(X)|r+\frac{z}{2} \rangle = \int dp\; e^{-ip\cdot z/\hbar}
\rho_m({\mathcal X}),
\end{equation}
the evaluation of $\rho^{\lambda \lambda'}_W(X) =\langle m_{\lambda}|\hat{\rho}_{m}(X)|m_{\lambda'}\rangle$ leads to
\begin{eqnarray}\label{eq:rho-to-map}
\rho^{\lambda \lambda'}_W(X) &=& (2 \pi \hbar)^N \int dx \; g_{\lambda \lambda'}(x) \rho_m({\mathcal X})\nonumber \\
&=&   (2 \pi \hbar)^N \int dx \; g_{\lambda \lambda'}(x) \rho_m^{{\mathcal P}}({\mathcal X}).
\end{eqnarray}

These relations allow one to transform operators expressed in the subsystem basis to Wigner representations of operators in the basis of
mapping states. The projected forms of the mapping operators and densities confine these quantities to the physical space and this
feature plays an important role in the discussions of the nature of dynamics using the mapping basis. We now show how these relations enter the expressions for expectation values and evolution equations.

\subsection{Forms of Operators in the Mapping Subspace}\label{subB}
We first consider the equivalent forms that operators take, provided they are confined to the physical mapping space. Since
\begin{eqnarray}\label{eq:identity}
\langle m_\lambda|\sum_\nu \hat{a}^\dagger_\nu  \hat{a}_\nu  |m_{\lambda'} \rangle
= \langle m_\lambda|m_{\lambda'} \rangle,
\end{eqnarray}
$\sum_\nu \hat{a}^\dagger_\nu \hat{a}_\nu$ is an identity operator in the mapping space.
(Here we include the explicit summation on mapping states for clarity.)
Using the definition of $g_{\lambda \lambda'}(x)$ in Eq.~(\ref{eq:g-def}), we may write the right side of Eq.~(\ref{eq:identity}) as
\begin{equation}
\langle m_\lambda|m_{\lambda'} \rangle=\int dx \; g_{\lambda \lambda'}(x).
\end{equation}
The left side of may be evaluated by inserting complete sets of coordinate states and taking
Wigner transforms so that an equivalent form for Eq.~(\ref{eq:identity}) is
\begin{eqnarray}
\int dx \; g_{\lambda \lambda'}(x) \sum_\nu c_{\nu \nu}(x)
 =\int dx \; g_{\lambda \lambda'}(x).
\end{eqnarray}
Thus, we see that
\begin{equation} \label{eq:sum-rule}
\sum_\nu c_{\nu \nu}(x)=\frac{1}{2\hbar} \sum_\nu (r_\nu^2 +p_\nu^2 -\hbar)=1,
\end{equation}
provided it lies inside the $g_{\lambda \lambda'}(x)$ integral.

This result has implications for the form of operators in the mapping basis. The matrix elements of an operator
$\hat{B}_W(X)$ in the subsystem basis may always be written as a sum of trace and traceless contributions,
\begin{equation}
B_{W}^{\lambda\lambda'}(X)= \delta_{\lambda \lambda'}({\rm Tr} B_{W})/N
+\overline{B}^{\lambda \lambda'}_W(X),
\end{equation}
where $\overline{B}^{\lambda \lambda'}_W(X)$ is traceless.  Inserting this expression into Eq.~(\ref{Bm_sub_wigner})
for $B_m({\mathcal X})$, we obtain
\begin{equation}
B_m({\mathcal X})=({\rm Tr} \;B_{W})/N +
\overline{B}_{W}^{\lambda\lambda'}(X)\overline{c}_{\lambda \lambda'}(x),
\end{equation}
provided $B_m({\mathcal X})$ appears inside the $g_{\lambda \lambda'}(x)$ integral. Note that all subsystem matrix elements
are of this form in view of Eq.~(\ref{eq:Bsubing}).  Here
$\overline{c}_{\lambda \lambda'}(x)=\frac{1}{2\hbar}[r_{\lambda}r_{\lambda'}
 +p_{\lambda}p_{\lambda'}+i(r_\lambda p_{\lambda'}-r_{\lambda'} p_\lambda)]$ is the traceless form of ${c}_{\lambda \lambda'}(x)$.

As a special case of these results, we can write the mapping Hamiltonian,
$H_m({\mathcal X})=H_{W}^{\lambda\lambda'}(X)c_{\lambda \lambda'}(x)$ in a convenient form.
The Hamiltonian matrix elements are given by
\begin{eqnarray}
{H}_W^{\lambda \lambda'}(X)&=&H_e(X) \delta_{\lambda \lambda'}+ \epsilon_\lambda \delta_{\lambda \lambda'}
+V_c^{\lambda \lambda'}(R)\nonumber \\
&\equiv& H_e(X) \delta_{\lambda \lambda'}+h^{\lambda \lambda'}(R),
\end{eqnarray}
which can be written as a sum of trace and traceless contributions,
\begin{eqnarray}
{H}_W^{\lambda \lambda'}(X)
&=& \Big(H_e(X) + ({\rm Tr}\; h)/N \Big)\delta_{\lambda \lambda'}
+\overline{h}^{\lambda \lambda'}(R)\nonumber \\
&\equiv& H_0(X)\delta_{\lambda \lambda'} +\overline{h}^{\lambda \lambda'}(R).
\end{eqnarray}
The Hamiltonian $H_0$ can be written as $H_0 \equiv P^2/2M+V_0(R)$. From this form for ${H}_W^{\lambda \lambda'}$, it follows that
\begin{equation} \label{eq:map-H-trace}
H_{m}({\mathcal X})=\frac{P^2}{2M}+V_0(R) +\frac{1}{2\hbar} \overline{h}^{\lambda\lambda'}(R)(r_{\lambda}r_{\lambda'}
 +p_{\lambda}p_{\lambda'}),
\end{equation}
again, when it appears inside integrals with $g_{\lambda \lambda'}(x)$. We have used the fact that $\overline{h}^{\lambda\lambda'}$ is
symmetric to simplify the expression for $\overline{c}_{\lambda \lambda'}(x)$ in this expression. This form of the mapping Hamiltonian
will play a role in the subsequent discussion.

\subsection{Expectation values}
Our interest is in the computation of average values of observables, such
as electronic state populations or coherence, as a function of time. The expression for the expectation value of a
general observable $\hat{B}_W(X)$ is
\begin{eqnarray}\label{eq:Bsub}
&&\overline{B(t)}=\int dX\; {\rm Tr} \; (\hat{B}_{W}(X)\hat{\rho}_{W}(X,t))=\\
&&\int dX\; B_{W}^{\lambda\lambda'}(X)\rho_{W}^{\lambda'\lambda}(X,t)
= \int dX\; B_{W}^{\lambda\lambda'}(X,t)\rho_{W}^{\lambda'\lambda}(X),\nonumber
\end{eqnarray}
where the trace is taken in the quantum subsystem space. In the
last line the time dependence has been moved from the density matrix to the operator, which also satisfies the
QCLE.

The expression for the expectation value can be written in the mapping basis using the results in the previous subsection. For
example, using Eq.~(\ref{eq:Bsubing}) and the first line of Eq.~(\ref{eq:rho-to-map}) we find
\begin{eqnarray}\label{eq:Bmapex}
&&\overline{B(t)}= \int dX\; \Big[ \int dx \; B_m({\mathcal X},t) g_{\lambda \lambda'}(x) \Big]  \\
&&  \qquad \times \Big[(2 \pi \hbar)^N \int dx' \; g_{\lambda' \lambda}(x') \rho_m(x',X) \Big] \nonumber \\
&&=\int d{\mathcal X} \; B_m({\mathcal X},t) \rho_{m}^{\mathcal P}({\mathcal X}) =
\int d{\mathcal X} \; B_m^{\mathcal P}({\mathcal X},t) \rho_{m}({\mathcal X}),\nonumber
\end{eqnarray}
where we have made use of the definition of the projection operator in Eq.~(\ref{eq:proj-op}) in writing the second equality.
The projection operator can instead be applied to the observable in view of the symmetry in the expression and the resulting form
is given in the last equality. We may write other equivalent forms for the expectation value. Starting from the second equality in
Eq.~(\ref{eq:Bsub}) involving the time evolved density and the time independent operator, we obtain
\begin{eqnarray}\label{eq:Bmapex2}
\overline{B(t)}&=&\int d{\mathcal X} \; B_m({\mathcal X}) \rho_{m}^{\mathcal P}({\mathcal X},t) \nonumber \\
&=&\int d{\mathcal X} \; B_m^{\mathcal P}({\mathcal X}) \rho_{m}({\mathcal X},t).
\end{eqnarray}
From a computational point of view, the penultimate equality in Eq.~(\ref{eq:Bmapex}) is most convenient since its evaluation entails
sampling from the initial value of the projected density and time evolution of the operator.

\subsection{Equations of motion}
The most convenient form of the expectation value requires a knowledge of
$B_m({\mathcal X},t)=B_{W}^{\lambda\lambda'}(X,t)c_{\lambda \lambda'}(x)$. Of course, if the solution to the QCLE in the
subsystem basis, $B_{W}^{\lambda\lambda'}(X,t)$, is known, this definition can be used directly to construct $B_m({\mathcal X},t)$;
however, the utility of the mapping basis representation lies in the fact that one can construct and solve the equation of motion for
$B_m({\mathcal X},t)$ directly. The derivation of the evolution equation was given earlier.~\cite{kim-map08} Here, we derive the evolution
equations by taking account of the properties of mapping operators under integrals of $g_{\lambda \lambda'}(x)$ in order to make
connection with the projected forms of operators and densities. This will allow us to explore the domain of validity
of the resulting equations.

The QCLE for an observable is expressed in the subsystem basis by taking matrix elements of the abstract equation $d\hat{B}_W(t)/dt=i\hat{{\mathcal L}} \hat{B}_W(t)$ with $i\hat{{\mathcal L}}$ defined in Eq.~(\ref{eq:qclop-abs}):
\begin{eqnarray}
&&\frac{d }{d t} \langle \lambda|\hat{B}_W (X, t)|\lambda' \rangle =
-\frac{i}{\hbar}\langle \lambda|[\hat{H}_W,\hat{B}_W (X, t)]|\lambda' \rangle \\
&& \quad + \frac{1}{2}\langle \lambda|(\{\hat{H}_W,\hat{B}_W (X, t)\}
-\{\hat{B}_W (X, t),\hat{H}_W\})|\lambda' \rangle. \nonumber
\end{eqnarray}
We may write this equation in terms of mapping variables using Eq.~(\ref{eq:Bsubing}) as
\begin{eqnarray} \label{eq:proj-eqmotmap1}
&&\int dx \; g_{\lambda \lambda'}(x) \frac{d }{d t}  B_m({\mathcal X},t) = \\
&& \int dx \; g_{\lambda \lambda'}(x) \Big(
-\frac{i}{\hbar}([\hat{H}_W,\hat{B}_W (X, t)])_m({\mathcal X},t) \nonumber \\
&& \quad + \frac{1}{2} (\{\hat{H}_W,\hat{B}_W (X, t)\}
-\{\hat{B}_W (X, t),\hat{H}_W\})_m({\mathcal X},t) \Big). \nonumber
\end{eqnarray}
The mapping variables occur inside integrals of $g_{\lambda \lambda'}(x)$ integral; i.e.,
they are projected onto the space of mapping states.
Since the commutator and Poisson bracket terms in this equation involve products of operators, we must obtain
the mapping form for a product of operators $\hat{A}_W(X)\hat{B}_W(X)$. The most direct way to make this
transformation is to consider the product of operators as they appear in the subsystem basis and then use
Eq.~(\ref{eq:Bsubing}) for each matrix element:
\begin{eqnarray}\label{eq:prod-1stform}
A_{W}^{\lambda \nu}(X)B_{W}^{\nu \lambda'}(X) &=& \int dx \; A_m(x,X) g_{\lambda \nu}(x) \\
&& \times \int dx' \; g_{\nu \lambda'}(x') B_m(x',X).  \nonumber
\end{eqnarray}
This expression does not lead to a useful form for the equations of motion. Instead we may write
\begin{eqnarray}
&&A_{W}^{\lambda \nu}(X) B_{W}^{\nu \lambda'}(X) = \langle \lambda|\hat{A}_W(X) \hat{B}_W (X)|\lambda' \rangle \nonumber \\
&& \qquad = \langle m_\lambda|\hat{A}_m(X) \hat{B}_m (X)|m_{\lambda'} \rangle  \\
&& \qquad  = \int dx \; g_{\lambda \lambda'}(x)  (\hat{A}_m(X) \hat{B}_m (X))_W({\mathcal X}) \nonumber
\end{eqnarray}
Given that the Wigner transform of a product of operators is
\begin{equation}
(\hat{A}_{m}(X)\hat{B}_{m}(X))_{W}={A}_{m}(x,X)e^{\hbar\Lambda_{m}/2i}{B}_{m}(x,X),
\end{equation}
where $\Lambda_{m}=\overleftarrow{\nabla_{p}}\cdot\overrightarrow{\nabla_{r}}
-\overleftarrow{\nabla_{r}}\cdot\overrightarrow{\nabla_{p}}$ is the negative of the Poisson bracket
operator on the mapping phase space coordinates, we obtain
\begin{eqnarray} \label{eq:prod-2ndform}
&&A_{W}^{\lambda \nu}(X)B_{W}^{\nu \lambda'}(X) = \\
&& \qquad \qquad \int dx \;  g_{\lambda \lambda'}(x) \Big(A_m({\mathcal X}) e^{\hbar\Lambda_{m}/2i}  B_m({\mathcal X})\Big).  \nonumber
\end{eqnarray}
In Appendix A we establish the equality between this form for the matrix product and that given in Eq.~(\ref{eq:prod-1stform}).
Inserting this result into Eq.~(\ref{eq:proj-eqmotmap1}), expanding the exponential operator and noting that the
mapping Hamiltonian is a quadratic function of the mapping phase space coordinates, we obtain (details of the
derivation are given in Ref.~[\onlinecite{kim-map08}])
\begin{equation}
\int dx \; g_{\lambda \lambda'}(x) \Big( \frac{d }{d t}  B_m({\mathcal X},t)
= i{\mathcal L}_m B_m({\mathcal X},t) \Big),
\end{equation}
where the mapping quantum-classical Liouville (MQCL) operator is given by the sum of two contributions:
\begin{eqnarray} \label{eq:QCLMop2}
 i{\mathcal{L}}_{m}=i{\mathcal{L}}_{m}^{PB} + i{\mathcal{L}}_{m}^{\prime}.
\end{eqnarray}
The Liouville operator $i{\mathcal{L}}_{m}^{PB}$ has a Poisson bracket form,
\begin{eqnarray} \label{eq:QCLMop}
&&i{\mathcal{L}}_{m}^{PB}= -\{H_{m}, \; \;\}_{{\mathcal X}}
 =\frac{\overline{h}^{\lambda\lambda'}}{\hbar}\left(p_{\lambda'} \frac{\partial}{\partial r_{\lambda}}-r_{\lambda'}
\frac{\partial}{\partial p_{\lambda}} \right) \nonumber \\
&&  \qquad \quad - \Big(\frac{\partial H_m}{\partial R}\cdot  \frac{\partial }{\partial P}
-\frac{P}{M}\cdot \frac{\partial }{\partial R} \Big),
\end{eqnarray}
where $\{\cdot,\cdot\}_{{\mathcal X}}$ denotes a Poisson bracket in the full mapping-environment phase space of the system,
while
\begin{eqnarray} \label{eq:QCLMprime}
 i{\mathcal{L}}_{m}^{\prime}=\frac{\hbar}{8}  \frac{\partial h^{\lambda\lambda'}}{\partial R}
 \Big(\frac{\partial^2}{\partial r_{\lambda'} \partial r_{\lambda}}
 +\frac{\partial^2}{\partial p_{\lambda'} \partial p_{\lambda}}\Big)
 \cdot\frac{\partial}{\partial P}.
\end{eqnarray}
In writing this form of the mapping Liouville operator we used the expression for the Hamiltonian given in Eq.~(\ref{eq:map-H-trace}).
This is allowed since by Eq.~(\ref{eq:prod-1stform}) the operators appear inside $g_{\lambda \lambda'}$ integrals.

The formal solution of the equation of motion for $B_m({\mathcal X},t)$ is
$B_m({\mathcal X},t)=e^{i{\mathcal L}_m t}B_m({\mathcal X})$. The expectation value of this operator is
given by (see Eq.~(\ref{eq:Bmapex}))
\begin{eqnarray}\label{eq:Bmapex3}
&&\overline{B(t)}= \int d{\mathcal X} \; \Big(e^{i{\mathcal L}_m t}B_m({\mathcal X})\Big) \rho_{m}^{\mathcal P}({\mathcal X})= \\
&& \int d{\mathcal X} \; B_m({\mathcal X}) e^{-i{\mathcal L}_m t}\rho_{m}^{\mathcal P}({\mathcal X}) \equiv \int d{\mathcal X} \;
B_m({\mathcal X}) \rho_{m}^{\mathcal P}({\mathcal X},t), \nonumber
\end{eqnarray}
where the evolution operator has been moved to act on the projected density using integration by parts. Thus,
we see that the projected density satisfies
\begin{eqnarray} \label{eq:qcl-rho-proj}
\frac{\partial}{\partial t} \rho_m^{{\mathcal P}}({\mathcal X},t) = -i{\mathcal{L}}_{m}\rho_{m}^{{\mathcal P}}(t).
\end{eqnarray}

Making use of the above results, we can establish relations among the various forms of the expectation values and the dynamics projected onto
the physical mapping states. From Eqs.~(\ref{eq:Bmapex2}) and (\ref{eq:Bmapex3}) we have the relation
$\int d{\mathcal X} \; B_m({\mathcal X}) \rho_{m}^{\mathcal P}({\mathcal X},t)=
\int d{\mathcal X} \; B_m^{\mathcal P}({\mathcal X}) \rho_{m}({\mathcal X},t)$. Differentiating both sides with respect to time and using the
MQCLE we may write this equality as
\begin{eqnarray}\label{eq:LPcommute}
&&\int d{\mathcal X} \; B_m({\mathcal X}) i{\mathcal L}_m {\mathcal P} \rho_{m}({\mathcal X},t)  \\
&& \qquad \qquad =\int d{\mathcal X} \; B_m({\mathcal X}) {\mathcal P} i{\mathcal L}_m \rho_{m}({\mathcal X},t). \nonumber
\end{eqnarray}
This identity, which is confirmed by direct calculation using the explicit form of $i{\mathcal L}_m$ in Appendix B,
shows that $i{\mathcal L}_m$ commutes with the projection operator. Thus, evolution under
the MQCL operator is confined to the physical mapping space.

\section{Trajectory Description of Dynamics} \label{sec:traj}
A variety of simulation schemes have been constructed for the solution of the QCLE, some involving trajectory based solutions~\cite{cecamchap,martens96,donoso98,santer01,sergi03b,mackernan08,wan00,wan02,horenko02,hanna05,hanna08}. These schemes involve either ensembles of surface-hopping trajectories or correlations among the trajectories. A solution in terms of an ensemble of independent trajectories evolving by Netwonian-like equations is not possible~\cite{kapral99}.

\subsection{Ensemble of entangled trajectories}
A trajectory based solution of the MQCLE can also be constructed but the trajectories comprising the ensemble are not independent. Such entangled trajectory solutions have been discussed by Donoso, Zheng and Martens~\cite{donoso01,donoso03} for the Wigner transformed quantum Liouville equation. While our starting equation is very different, a similar strategy can be used to derive a set of equations of motion for an ensemble of entangled trajectories.

The MQCLE (\ref{eq:qcl-rho}) can be written as a continuity equation in the full (mapping plus environment) phase space as
\begin{eqnarray} \label{eq:continuity}
\frac{\partial}{\partial t}\rho_{m}^{{\mathcal P}}({\mathcal X},t)&=&-\frac{\partial}{\partial {\mathcal X}} \cdot j({\mathcal X},t)\\
&=& -\frac{\partial}{\partial {\mathcal X}} \cdot [v({\mathcal X};\rho_{m}^{{\mathcal P}}({\mathcal X},t))\rho_{m}^{{\mathcal P}}({\mathcal X},t)] \nonumber  ,
\end{eqnarray}
where the current $j({\mathcal X})=(j_r,j_p,j_R,j_P)$ has components:
\begin{eqnarray}
&&j_{r_{\lambda'}}= \frac{\overline{h}^{\lambda\lambda'}}{\hbar} p_\lambda \rho_{m}^{{\mathcal P}}, \;
j_{p_{\lambda'}}= -\frac{\overline{h}^{\lambda\lambda'}}{\hbar} r_\lambda \rho_{m}^{{\mathcal P}}, \;
j_R=\frac{P}{M} \rho_{m}^{{\mathcal P}},\\
&&j_P = -\frac{\partial H_{m}}{\partial R} \rho_{m}^{{\mathcal P}} +\frac{\hbar}{8}\frac{\partial \overline{h}^{\lambda\lambda'}}{\partial R}\Big(\frac{\partial^2}{\partial r_{\lambda'} \partial r_{\lambda}}+\frac{\partial^2}{\partial p_{\lambda'} \partial p_{\lambda}}\Big) \rho_{m}^{{\mathcal P}}. \nonumber
\end{eqnarray}
The second equality in Eq.~(\ref{eq:continuity}) defines the phase space velocity field $v({\mathcal X};\rho_{m}^{{\mathcal P}}({\mathcal X},t))$ through $j({\mathcal X},t)\equiv v({\mathcal X};\rho_{m}^{{\mathcal P}}({\mathcal X},t))\rho_{m}^{{\mathcal P}}({\mathcal X},t))$, which is a functional of the full phase space density.

We seek a solution in terms of an ensemble of ${\mathcal N}$ trajectories, $\rho_m^{{\mathcal P}}({\mathcal X},t)={\mathcal N}^{-1}\sum_{i=1}^{{\mathcal N}} {\rm w}_i \delta({\mathcal X}-{\mathcal X}_i(t))$, where ${\rm w}_i$ is the initial weight of trajectory $i$ in the ensemble.
To find the equations of motion for the trajectories, consider the phase space average of the product of an arbitrary function $f({\mathcal X})$ with Eq.~(\ref{eq:continuity}):
\begin{eqnarray} \label{eq:Bcont}
&&\frac{d}{d t}\int d{\mathcal X}\; f({\mathcal X}) \rho_{m}^{{\mathcal P}}({\mathcal X},t)=\\
&&\qquad \qquad \int d{\mathcal X}\; \frac{\partial f({\mathcal X})}{\partial {\mathcal X}} \cdot [v({\mathcal X};\rho_{m}^{{\mathcal P}}({\mathcal X},t))\rho_{m}^{{\mathcal P}}({\mathcal X},t)] \nonumber  ,
\end{eqnarray}
where we have carried out an integration by parts to obtain the right side of the equality. Substitution of the ansatz for the phase space density into this equation gives
\begin{eqnarray}
\sum_{i=1}^{{\mathcal N}} {\rm w}_i \frac{\partial f({\mathcal X}_i(t))}{\partial {\mathcal X}_i(t)} \cdot \Big[ \dot{{\mathcal X}}_i(t)- v({\mathcal X}_i(t);\rho_{m}^{{\mathcal P}}({\mathcal X}_i(t)))\Big]=0,
\end{eqnarray}
from which it follows that the trajectories satisfy the evolution equations, $\dot{{\mathcal X}}_i(t)= v({\mathcal X}_i(t);\rho_{m}^{{\mathcal P}}({\mathcal X}_i(t)))$. More explicitly we have
\begin{eqnarray} \label{eq:entangled}
\dot{r}_{\lambda} &=& \frac{\partial H_m}{\partial p_\lambda}, \quad
\dot{p}_{\lambda} =-\frac{\partial H_m}{\partial r_{\lambda}},  \quad
\dot{R} = \frac{\partial H_m}{\partial  P},\\
\dot{P} &=& -\frac{\partial H_{m}}{\partial R}
+\frac{\hbar}{8 \rho_{m}^{{\mathcal P}}}\frac{\partial \overline{h}^{\lambda\lambda'}}{\partial R}\Big(\frac{\partial^2}{\partial r_{\lambda'} \partial r_{\lambda}}+\frac{\partial^2}{\partial p_{\lambda'} \partial p_{\lambda}}\Big) \rho_{m}^{{\mathcal P}}. \nonumber
\end{eqnarray}
The second term in the environmental momentum equation couples the dynamics of all members of the ensemble since it involves the phase space density.

\subsection{Ensemble of independent trajectories}

If the last term in the $\dot{P}$ equation is dropped we recover simple Newtonian evolution equations:
\begin{eqnarray}
\label{Lm-gen}
\frac{d r_{\lambda}}{dt}&=& \frac{\partial H_m}{\partial p_\lambda}, \qquad
\frac{d p_{\lambda}}{dt}=-\frac{\partial H_m}{\partial r_{\lambda}}, \\
\frac{d R}{dt} &=& \frac{\partial H_m}{\partial  P}, \qquad
\frac{d P}{dt}= -\frac{\partial H_m}{\partial R} .\nonumber
\end{eqnarray}
This result also follows from the fact that neglect of the last term in the $\dot{P}$ equation corresponds to the neglect of the
last term in the formula for $i{\mathcal L}_m$ in Eq.~(\ref{eq:QCLMop2}). Thus, in this approximation
\begin{equation}\label{qclm}
\frac{\partial}{\partial t}\rho^{{\mathcal P}}_m ({\mathcal X},t) = \big\{ H_m, \rho^{{\mathcal P}}_m \big\}_{{\mathcal X}} \equiv
-i \mathcal{L}_m^{PB} \rho^{{\mathcal P}}_m ({\mathcal X},t),
\end{equation}
which we call the Poisson bracket mapping equation (PBME). Since the approximate evolution has a Poisson bracket
form, it admits a solution in characteristics and the corresponding ordinary differential equations are those above
in Eq.~(\ref{Lm-gen})~\cite{kim-map08}.

In contrast to Eq.~(\ref{eq:LPcommute}), in Appendix B we show that
\begin{eqnarray}
&&\int d{\mathcal X}\; B_m({\mathcal X})i \mathcal{L}_m^{PB} {{\mathcal P}} \rho_m({\mathcal X}) \\
&& \qquad \quad \ne \int d{\mathcal X}\; B_m({\mathcal X}){{\mathcal P}} i \mathcal{L}_m^{PB} \rho_m({\mathcal X}).
\nonumber
\end{eqnarray}
Consequently,  the Poisson bracket mapping operator $i \mathcal{L}_m^{PB}$ does not commute with the projection operator. Therefore,
unlike the evolution under the full MQCL operator, the evolution prescribed by the PBM operator may take the dynamics out of the physical space.

As will be seen shortly, one consequence of the dynamics leaving the physically relevant regions of phase space is a lack of stability
of trajectories due to inversion of the potential for bath coordinates. It is therefore important to minimize artificial instabilities arising due to the use of too large a time step in numerical methods of solving the evolution equations.  We note that as in the case of Brownian motion, the bath coordinates typically evolve on a much longer time scale than the subsystem phase space coordinates, as can be seen from a scaling analysis of the equations of motion Eq.~(\ref{Lm-gen}) in terms of the dimensionless mass ratio $\epsilon = (m/M)^{1/2}$. As a consequence, one might expect that the motion of the subsystem limits the size of the time step utilized in the integration scheme, and small time steps must be chosen to deal with regions of phase space in which rapid changes in population occur.  In Appendix C we show that an integrator may be designed using the exact solution of the subsystem equations of motion when the bath position is held fixed.  Using this integrator, numerical instabilities are minimized, allowing us to focus on true instabilities inherent in the physical system arising from the PBME approximation.

We also remark that although these equations of motion have been derived from an approximation to QCL dynamics in the mapping basis, they also appear in the in the semi-classical path integral investigations of quantum dynamics by Stock and Thoss~\cite{thoss99,stock05} and in the linearized semiclassical-initial value representation (LSC-IVR) of Miller~\cite{miller78,miller01,miller07}. These results indicate that LSC-IVR dynamics is closely related to this approximate form of the QCLE. Connections between QCL dynamics and linearized path integral formulations have been discussed in the literature~\cite{shi04a,bonella10}. The utility of this approximation to the QCLE hinges on the form of the Hamiltonian and the manner
in which expectation values are computed. These issues are also discussed in the next section.

\section{Dynamical instabilities in approximate evolution equations}\label{sec:PBME}

In Sec.~\ref{subB} we showed that the mapping Hamiltonian,
\begin{eqnarray}\label{eq:map-H}
&&H_m({\mathcal X})=H^{\lambda \lambda'}_W(X) c_{\lambda \lambda'}(x)=\\
&&H^{\lambda \lambda'}_W(X)\frac{1}{2\hbar}
[r_\lambda r_{\lambda'}+ p_{\lambda}p_{\lambda'} + i(r_\lambda p_{\lambda'} - r_{\lambda'} p_{\lambda})
- \hbar\delta_{\lambda\lambda'}], \nonumber
\end{eqnarray}
could be written in the equivalent form given in Eq.~(\ref{eq:map-H-trace}), provided the Hamiltonian operator appears inside
the $g_{\lambda \lambda'}$ integral; i.e.,
is projected onto the physical space. In view of Eq.~(\ref{eq:prod-1stform}), and its equivalence to Eq.~(\ref{eq:prod-2ndform}), this condition is
satisfied for evolution under the MQCLE. Evolution under MQCL dynamics is confined to the physical space and the two forms of the Hamiltonian will
yield equivalent results. In this section we discuss instabilities that may arise in approximations to the MQCL as a result of the dynamics
taking the system outside of the physical space. Problems associated with the lack of confinement to the physical space in other mapping
formulations have been discussed earlier by Thoss and Stock~\cite{thoss99}. Here we reconsider some aspects of these issues in the context
of the QCL formulation.

While different forms of the mapping Hamiltonian are equivalent in the mapping
subspace, should the dynamics take the system out of this space, the evolution
generated by the different Hamiltonian forms will not be the same. Indeed, depending on precise form of the dynamics, instabilities
can arise that depend on the form of the Hamiltonian that is employed. In particular, from the structure of $H_m$ in
Eq.~(\ref{eq:map-H}), one can see that it is possible encounter ``inverted" potentials
if the quantity in square brackets is negative. This problem has appeared in
approximate schemes based on the mapping formulation and suggestions for its partial remedy
have been suggested~\cite{stock05,bonella01,bonella05}. Such investigations have led to the observation that the form
of $H_m$ in Eq.~(\ref{eq:map-H-trace}), where the resolution of the identity is used
to simplify the Hamiltonian form, provides the best results.

Even if such inverted potentials are not present at the initial phase points
of the trajectories representing the evolution of the density matrix, they may still arise in the course of approximate evolution that may take the system
outside the physical space; for example, under PBME dynamics. To investigate the conditions under which unstable dynamics appear,
consider systems that have localized regions of
strong coupling among diabatic states and asymptotic regions where such coupling vanishes. The Hamiltonian
matrix is approximately diagonal in the asymptotic regions and in such regions $H_m$ takes the form,
\begin{equation} \label{Hasymp}
H_m \sim \frac{P^{2}}{2M} +
    V_0(R) +\sum_{\lambda}\overline{h}^{\lambda\lambda} \Gamma_\lambda \equiv  \frac{P^{2}}{2M} + V_{\rm asy},
\end{equation}
where we have defined
$\Gamma_\lambda = \frac{1}{2\hbar}(r_\lambda^2 + p_{\lambda}^2)$. The second equality
defines the effective asymptotic potential energy $V_{\rm asy}$. Since
$\{\Gamma_\lambda,H_m\}_{{\mathcal X}}=0$, the $\Gamma_\lambda$ are conserved in the asymptotic
regions and can be considered constants.

The effective asymptotic potential energy can be written in the equivalent form,
\begin{equation} \label{asympot}
  V_{\rm asy} =   V_0(R)     +\sum_{\lambda}h^{\lambda\lambda}\Delta \Gamma_\lambda ,
\end{equation}
where $\sum_\lambda \Gamma_\lambda=\Gamma$ and  $\Delta \Gamma_\lambda  =
\Gamma_\lambda - \Gamma/N$, which satisfies
$\sum_{\lambda}\Delta \Gamma_\lambda=0$.
From this equation we see that if the matrix element $h^{\lambda\lambda}$ dominates asymptotically,
an inverted potential will be possible if $\Delta \Gamma_\lambda < -\frac{1}{N}$. If instead $V_0$ dominates asymptotically,
no instability will occur. Likewise, if another $h^{\lambda'\lambda'}$ grows more quickly asymptotically,
and does not lead to an inverted potential contribution, it will compensate
for the inversion due to the $h^{\lambda\lambda}$ term. Note that not all
$h^{\lambda\lambda}$ terms can give rise to inverted contributions at the same
time because $\sum_{\lambda}\Delta \Gamma_\lambda=0$.  An interesting case occurs
when all $h^{\lambda\lambda}$ grow asymptotically in the same way, e.g. as $\tilde
h$. In that case, the asymptotic potential takes the form  $V_{\rm asy}= V_0(R)$,
which is never inverted.

Thus, if not all $h^{\lambda\lambda}$ have the
same asymptotic behavior and $V_0$ is not asymptotically dominant,
then it is possible that inverted potentials may occur. In these cases, even if the initial condition is such that
an inverted potential does not exist, as the system moves through the coupling
region and into the asymptotic region, one can encounter cases where
$\Delta \Gamma_\lambda < -\frac{1}{N}$, which may result in an inverted effective potential.

\subsection{Simulations of the dynamics}\label{sec:results}

While the evolution prescribed by the PBME in Eq.~(\ref{qclm}) may take the system outside the physical mapping space resulting in dynamical instabilities that could affect the quality of the solutions, simulations on a variety of systems has shown that often very accurate results can be obtained at a computational cost that is far less than that for simulations of the full QCLE. For example, accurate results for the spin-boson system~\cite{kim-map08}, simple curve crossing models~\cite{nassimi10} and the room temperature excitation transfer in the Fenna-Mathews-Olsen light harvesting complex~\cite{kelly-FMO} have been obtained using this method. In this section we have chosen examples to illustrate cases where the simulations of the PBME exhibit more serious deviations from the solutions of the full QCLE and exact quantum dynamics as a result of the effects discussed above.

\subsubsection{Curve crossing dynamics: nuclear momentum distributions}
The simple curve crossing model~\cite{tully90} with Hamiltonian
\begin{eqnarray}
&&H_{m}({\mathcal X})=P^2/2M + \overline{h}^{\lambda\lambda'}(R)(r_{\lambda}r_{\lambda'}  +p_{\lambda}p_{\lambda'}), \nonumber \\
&&\overline{h}^{11}=-\overline{h}^{22}=A[1 - e^{-B |R|}]R/|R|, \nonumber \\
&&\overline{h}^{12}=\overline{h}^{21}=Ce^{-DR^2},
\end{eqnarray}
is one of the common benchmark cases for quantum dynamics. In this model $H_W^{\lambda \lambda'}$ is traceless so the forms of the mapping Hamiltonian in Eqs.~(\ref{eq:map-H-trace}) and (\ref{eq:map-H}) are identical. Quantitatively accurate results for population transfer and coherence have been obtained for this system using PBME dynamics~\cite{nassimi10}, so we focus instead on the properties of the nuclear degrees of freedom.

We have shown~\cite{nassimi10} that only part of the back coupling of the quantum subsystem on the bath is accounted for in this formulation so that the evolution of the classical degrees of freedom may differ from that in the full QCLE. Simulations of this model system~\cite{miller07} using
LSC-IVR approximations to path integral dynamics have shown that the nuclear momentum
distribution, after the system passes through the avoided crossing, has single peak. More accurate simulations based on the
forward-backward (FB)-IVR yield a double-peak structure in accord with exact quantum results. As the system passes through the avoided crossing and the coupling vanishes, the nuclear momenta have characteristically different values in the two asymptotic states giving rise to a bimodal distribution. The single-peaked structure of the LSC-IVR simulations was attributed to the mean-field nature of the
nuclear dynamics in this approximation to the dynamics~\cite{miller07}.

\begin{figure}[p]       \centering
    \includegraphics[width=.8\columnwidth,angle=-90]{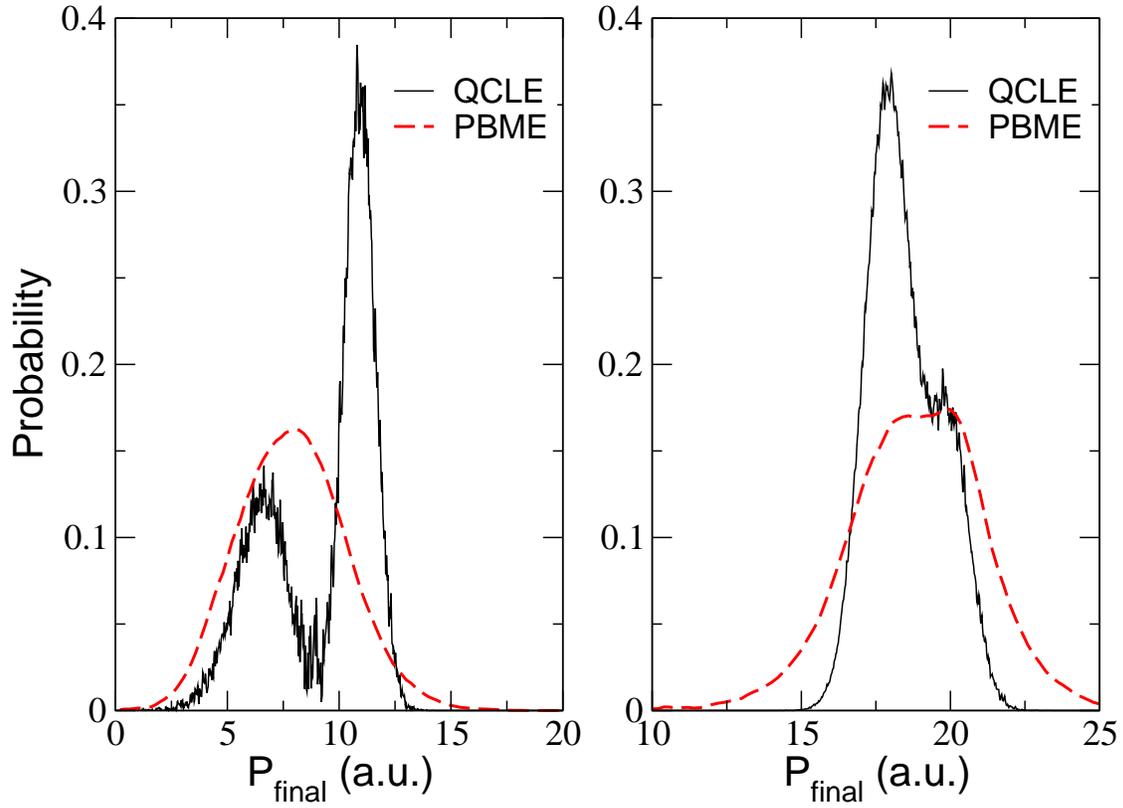}
    \caption{Plots of the momentum distribution $p(P_{{\rm final}})$ after passage through the avoided crossing: QCLE (solid lines), PBME (dashed lines).  The parameter values are $A=0.01$, $B=1.6$, $C=0.005$ and $D=1$ (both panels), and the initial momentum is $P_0=11$ (left panel) and $P_0=20$ (right panel). All parameters are reported in atomic units.
}
    \label{fig:avoid-cross}
\end{figure}

Here we present comparisons of the nuclear momentum distributions obtained from the simulations of the
QCLE using a Trotter-based algorithm~\cite{mackernan08} and its approximation by the PBME. We expect the PBME to yield results similar to those of  LSC-IVR since the evolution equations are similar in these approximations~\cite{footnote:PBsims}. The momentum distributions are shown in Fig.~\ref{fig:avoid-cross}.

The PBME simulations do indeed yield a momentum distribution with a single peak. The full QCLE simulations are able to reproduce the correct
double-peak structure of this distribution~\cite{footnote:two-peaks}, indicating that the failure of the PBME to capture this effect is due to the approximations made to obtain this evolution equation, and not the underlying QCL description.

\subsubsection{Conical Intersection Model}
A two-level, two-mode quantum model for the coupled vibronic states of a linear $ABA$ triatomic molecule has been constructed by Ferretti, Lami, and Villiani (FLV)~\cite{ferretti_1,ferretti_2} in their investigation of the dynamics near a conical intersection. The nuclei are described using two vibrational degrees of freedom: a symmetric stretch, $X$, the tuning coordinate and an anti-symmetric stretch coupling coordinate, $Y$. We denote the mapping Hamiltonian for this model by $H_{m}^s(R_s,P_s,x)$ whose form is given by Eq.~(\ref{eq:map-H-trace}) with
\begin{eqnarray} \label{feretti:hamiltonian}
 &&H_0(R_s,P_s) =\Big( \frac{P_X^2}{2 M_X} + \frac{P_Y^2}{2 M_Y}\Big)+ \frac{\Delta}{2} \\
 && \quad +\frac{1}{2}M_Y \omega_Y^2 Y^2 +\frac{1}{2}M_X \omega_X^2[(X-X_1)^2+(X-X_2)^2] ,\nonumber
\end{eqnarray}
and
\begin{eqnarray}
&&    \overline{h}^{11}=-\overline{h}^{22} = \frac{1}{2}M_X \omega_X^2\Big[X(X_2-X_1) +\frac{1}{2}(X_1^2-X_2^2)\Big],
\nonumber    \\
&&    \overline{h}^{12} =\overline{h}^{21}=  \gamma Y e^{-\alpha (X-X_3)^2} e^{-\beta Y^2},
\end{eqnarray}
In these equations $R_s=(X,Y)$ while $P_s=(P_X,P_Y)$, $(M_X, M_Y)$ and $(\omega_X, \omega_Y)$ are the momenta, masses and frequencies
of the $X$ and $Y$ degrees of freedom.~\cite{footnote-notation} If the FLV model is bilinearly coupled to a bath of independent harmonic
oscillators the Hamiltonian has the form,
\begin{eqnarray}
&& H_m(R,P) =  H_m^{s}(R_s,P_s,x) \\
&&\qquad + \sum_j^{N_B} \frac{P_j^2}{2M_j}+  \frac{M_j \omega_j^2}{2} (R_j - \frac{c_j}{M_j\omega_j^2} X)^2 \nonumber  \\
&&\qquad +  \sum_l^{N_B} \frac{P_l^2}{2M_l} + \frac{1}{2}M_l \omega_l^2 (R_l - \frac{c_l}{M_l\omega_l^2} Y)^2  .\nonumber
\end{eqnarray}
The coordinates and momenta of each bath oscillator with mass $M_j$ are $(R_{j}, P_j)$ and
and $N_B$ is the number of oscillators. The coupling constants and frequencies, $c_j$ and $\omega_j$, correspond with those of a
harmonic bath with an Ohmic spectral density. The dynamics of the FLV model, with and without coupling to the oscillator bath, was studied
in detail Ref.~[\onlinecite{kelly10}] using Trotter-based simulations of the QCLE. Below we present results on this model using the
approximate PBME dynamics.

\begin{figure}[p]       \centering
    \includegraphics[width=.8\columnwidth,angle=-90]{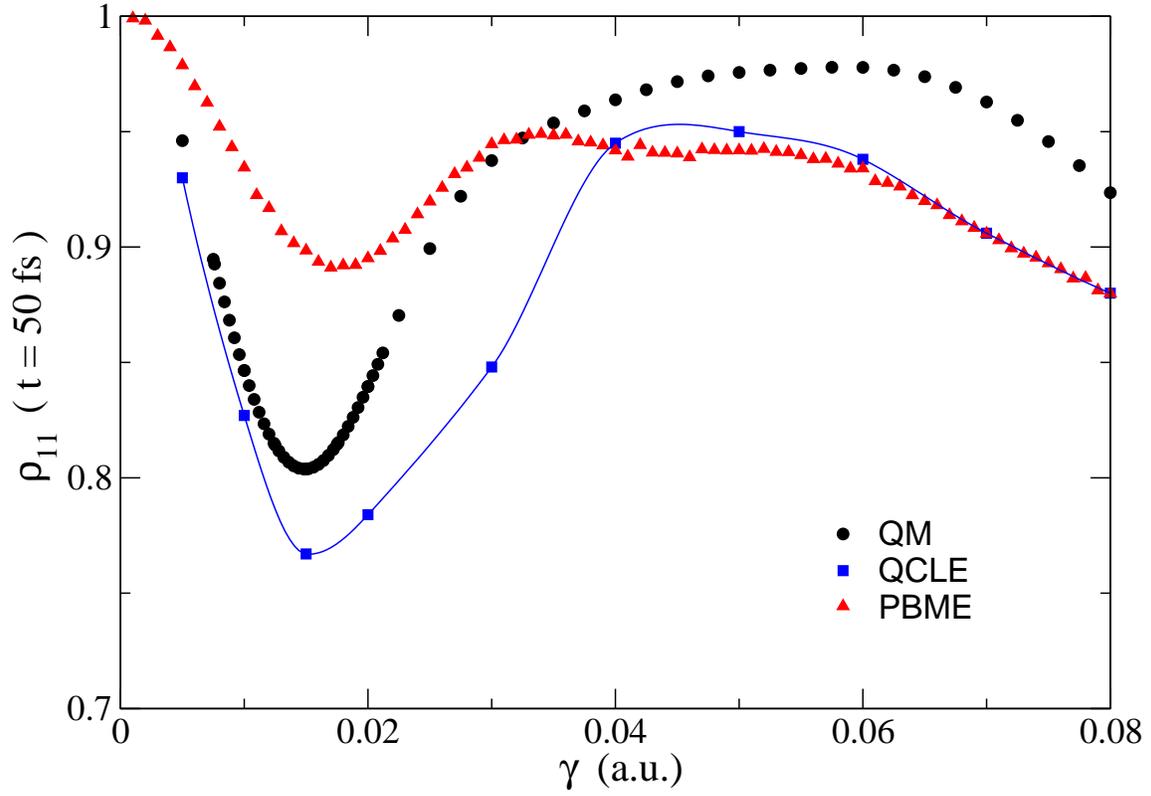}
    \caption{Ground adiabatic state populations $P_{S_1}(t=50\;{\rm fs})$ versus $\gamma$. The quantum results are taken from Ref.~[\onlinecite{ferretti_1}]
    and the QCLE results are from Ref.~[\onlinecite{kelly10}]. The parameters in the FLV model are: $\omega_X=0.001$, $\omega_Y=0.00387$,
    $M_X=20000$, $M_Y=6667$, $\alpha=3$, $\beta=1.5$, $X_1=4.$, $X_2=X_3=3.$ and $\Delta=0.01$, all in atomic units.}
    \label{fig:methods}
 \end{figure}

In Fig.~\ref{fig:methods} we compare the PBME results for ground adiabatic state populations $P_{S0}$ at $t = 50$ fs as a function of
the coupling strength, $\gamma$, with exact quantum and full QCLE results.

We see that general shape, including the appearance of a minimum and maximum in the probability as $\gamma$ increases, is
captured by all methods. The QCLE results reproduce the exact quantum results for the FLV population transfer curve in the low coupling
range and deviate somewhat for intermediate and high values of the coupling. The PBME results are less accurate at low coupling strengths and
match the QCLE simulations at high coupling. While not quantitatively accurate over the full coupling range, the PBME results capture the essential
physics in these curves.

It is interesting to examine statistical features of the ensemble of independent trajectories that were used to obtain these results. When the
calculation were carried out using the original mapping Hamiltonian with form in Eq.~(\ref{eq:map-H}) we found that $60\%$ of the ensemble was
initially on an inverted surface. Furthermore, $61\%$ of trajectories in the ensemble experienced an inverted surface at least one time step during the evolution and $50\%$ of the trajectories in the ensemble diverged. If instead the Hamiltonian with form in Eq.~(\ref{eq:map-H-trace}) was used
$0.1\%$ of the ensemble was initially on an inverted surface, $7\%$ of the ensemble experienced an inverted surface at least one time step
during the evolution and no trajectories in the ensemble diverged. In accord with other investigations, these results indicate the sensitivity
of the approximate evolution equations to the form of the mapping Hamiltonian. Many of the effects arising from instabilities can be
ameliorated by first separating the Hamiltonian matrix into trace and traceless parts and employing the resolution of the identity.

\begin{figure}[p]       \centering
 \includegraphics[width=.8\columnwidth,angle=-90]{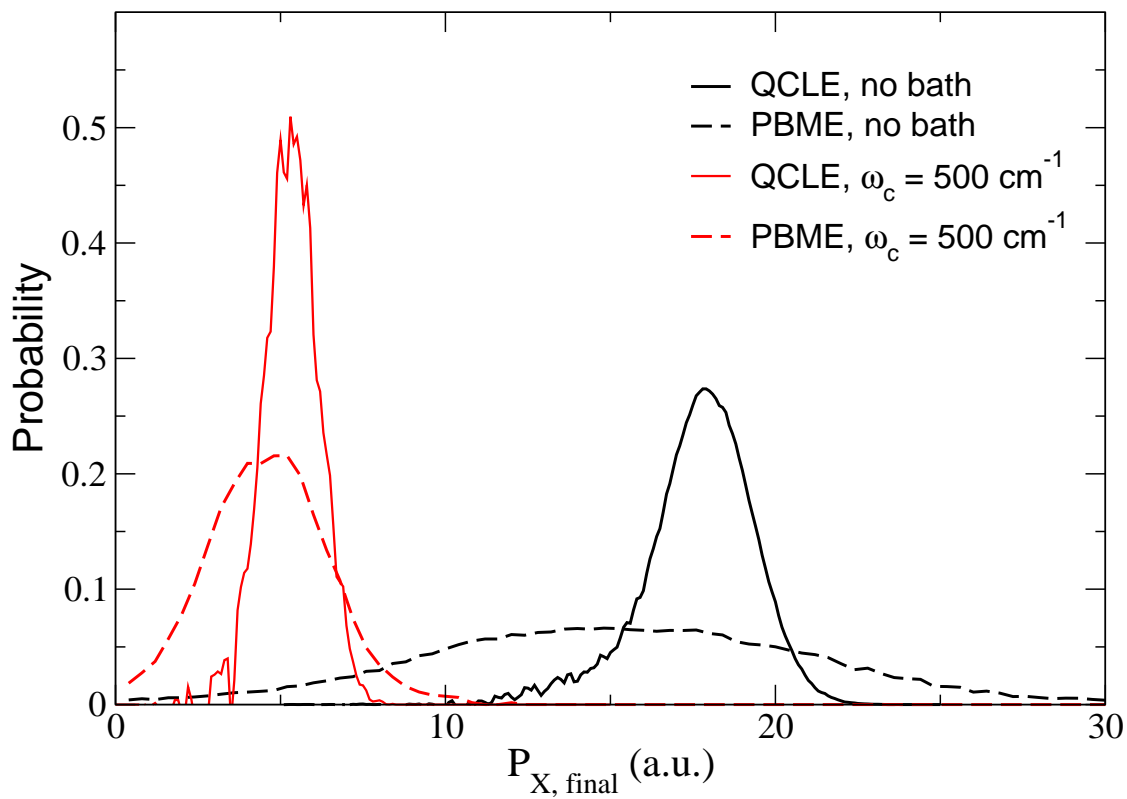}
        \caption{$P_X$ momentum distributions after passage through the conical intersection. The plot shows distributions obtained from
        simulations of the QCL and PBM equations for the FLV model without and with coupling to a bath of harmonic oscillators. The number of oscillators is $N_B=100$ and the temperature is $T=300 {\rm K}$.}
    \label{fig:FLV-momentum}
\end{figure}

Figure~\ref{fig:FLV-momentum} compares the $P_X$ momentum distributions of the FLV model after passage through the conical intersection obtained
from simulations of the full QCLE and its PBME approximation. The figure also presents results for this momentum distribution when the FLV model
is coupled to a harmonic bath. The QCLE distribution is much narrower than that obtained from the PBME simulations and the peak is shifted to
somewhat smaller momenta. This trend persists when a larger environment is present but the distributions are in much closer accord. This is
consistent with the fact that the PBME does not properly account for a portion of the influence of the quantum system on its environment. For
a larger many-body environment the effect of such back coupling will be smaller.

\subsubsection{Collinear Reactive Collision Model}

Finally, we consider a two-level, two-mode quantum model~\cite{martinez98} for the collinear triatomic
reaction $A+BC \to AB+C$. The diabatic states of the system are functions of $R=(\overline{X},\overline{Y})$, where $\overline{X}$ is
the distance between atoms B and C, while $\overline{Y}$ is the distance between atom A and the center of mass of the diatomic BC.
The mapping Hamiltonian again has the form given by Eq.~(\ref{eq:map-H-trace}) with
\begin{eqnarray} \label{reaction:hamiltonian}
 &&H_0(X,Y,P_X,P_Y) =\Big( \frac{P_X^2}{2 M_X} + \frac{P_Y^2}{2 M_Y}\Big) \\
 && \quad + \frac{D_e}{2}(1-e^{-\alpha(\overline{X}-\overline{X}_0)})^2 +\frac{D_{r}}{2}e^{-\alpha(\overline{Y}-\overline{X}/2-\overline{X}_0)}\nonumber \\
 && \quad  +\frac{D_e}{2}(1-e^{-\alpha(\overline{Y}-\overline{X}/2-\overline{X}_0)})^2+\frac{D_{r}}{2}e^{-\alpha(\overline{X}-\overline{X}_0)} \nonumber
\end{eqnarray}
and
\begin{eqnarray}
\overline{h}^{11}&=& \frac{D_e}{2}(1-e^{-\alpha(\overline{X}-\overline{X}_0)})^2 +\frac{D_{r}}{2}e^{-\alpha(\overline{Y}-\overline{X}/2-\overline{X}_0)}\nonumber \\
&-&\frac{D_e}{2}(1-e^{-\alpha(\overline{Y}-\overline{X}/2-\overline{X}_0)})^2-\frac{D_{r}}{2}e^{-\alpha(\overline{X}-\overline{X}_0)} \nonumber \\
\overline{h}^{12} &=&\overline{h}^{21}= \Delta,
\end{eqnarray}
with $\overline{h}^{11}=-\overline{h}^{22}$. In these equations $(P_X, P_Y)$ and $(M_X, M_Y)$ are the momenta and inertial
masses corresponding to the $BC$ and $A-BC$ degrees of freedom, respectively. This model describes two separate diabatic surfaces,
and the off-diagonal diabatic coupling matrix elements are constant.

The QCLE for this model has been simulated in the diabatic basis using the multiple spawning molecular dynamics method~\cite{martinez98}
and the results are in quantitative agreement with numerically exact quantum dynamics~\cite{wan02}. This system provides an interesting test
case since the dynamics can, in principle, explore unphysical regions in the model equations. Divergences occur where the (diagonal)
elements of Hamiltonian are large; i.e., for large negative values of $\overline{X}$ and $\overline{Y}-\overline{X}/2$.
While the model allows these negative values, physically, they represent distances which should not become negative and the model
loses its validity in these regions. Because the potential is large for large, negative values of these coordinates,
the nonphysical regions are exponentially suppressed if full quantum or full QCL simulations are
carried out and physically meaningful results can be obtained with this Hamiltonian. This is not the case for the dynamics
given by Eqs.~(\ref{Lm-gen}) due to the instability from the inverted potential, and these approximate evolution equations
are much more sensitive to the form of the potential.

In order to ensure that the coordinates of the system do not diverge, the model can be altered to avoid nonphysical values of the coordinates.
A reasonable adjustment of the model that keeps the values of $\overline{X}$ and $\overline{Y}$ bounded, even in the approximate
PBME, is to add a steep confining potential,
\begin{eqnarray}
  V_{a}(R)= D_e \left(e^{-z\alpha(\overline{X}-\tilde
  X_0)}+e^{-z\alpha(\overline{Y}-\overline{X}/2-\tilde X_0)}\right),
\label{Vecorr}
\end{eqnarray}
where $z$ and $\tilde X_0$ are parameters. We have chosen the following values: $z=4$ and $\tilde X_0= \overline{X}_0/2$.
By denoting the additional potential as $V_a$, the adjusted Hamiltonian is still of the same general form, so none of
the formalism needs to be changed. We confirmed that this added potential does not substantially change the physical problem~\cite{footnote:RM}.

\begin{figure}[p]
\includegraphics[width=.8\columnwidth,angle=-90]{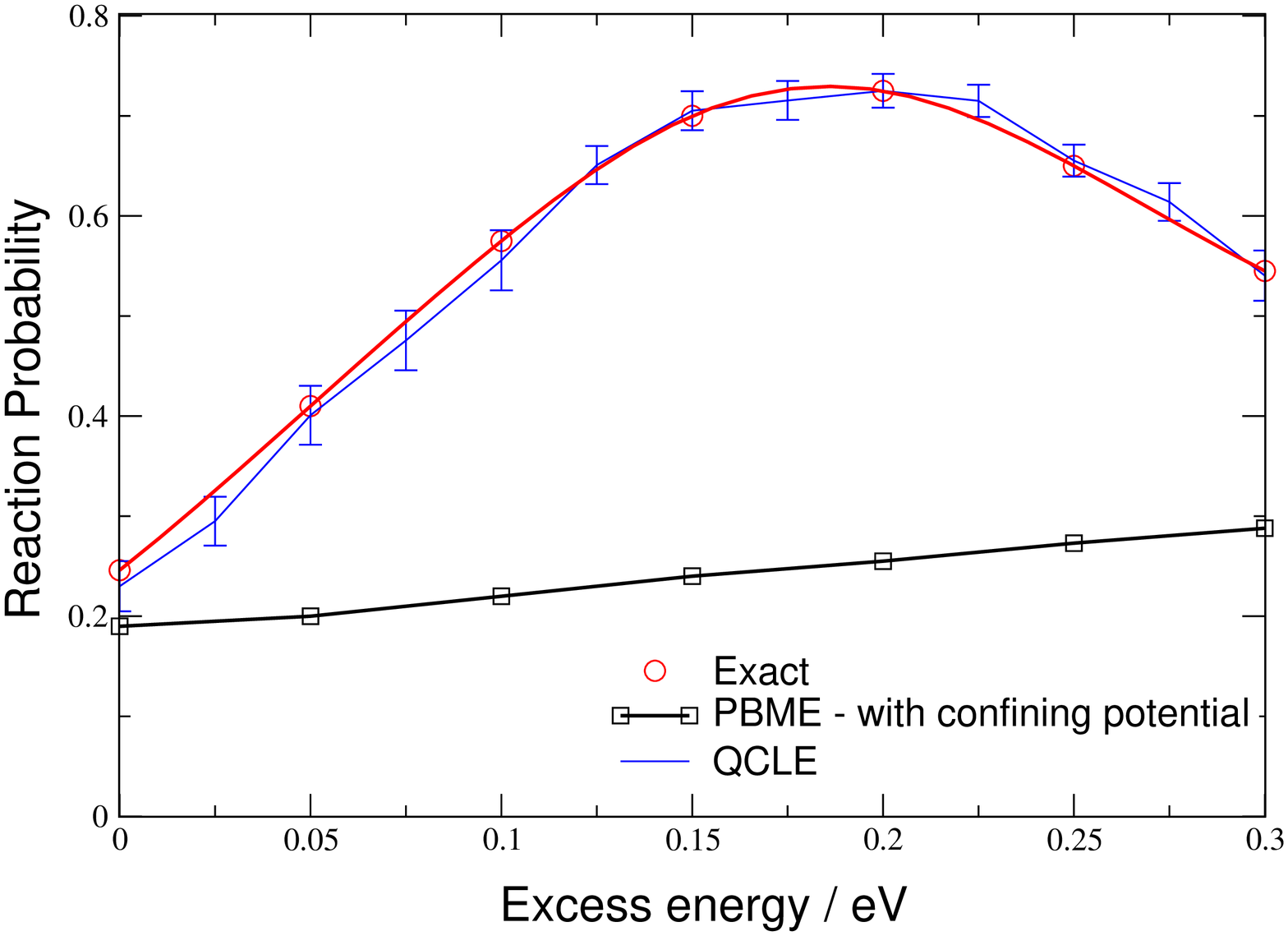}
 \caption{Comparison between the quantum-mechanical and full QCLE reaction
  reaction probabilities, with that given by the approximate PBME
  dynamics, as a function of the excess energy. Parameter values:
  $M_X = 6289,  M_Y = 8385, \Delta = 0.00136, \alpha = 0.458038, X_0 = 5.0494,
  D_e  = 0.038647, D_{rep}= 0.02$. (All quantities in atomic units.)
\label{fig:excess}}
\end{figure}

In the simulations of the reaction dynamics, the initial wave packet was directed towards the reaction region by giving it a non-zero
$Y$-momentum. This initial momentum can  be converted to an excess energy, which is roughly the kinetic energy minus the energy of
the barrier in the reaction. The results of simulations of the PBME are compared with exact quantum and full QCLE simulations
in Fig.~\ref{fig:excess}. This figure plots the reaction probability versus the excess energy. The approximate PBME dynamics
fails to capture the peaked structure of the reaction probability but does yield probabilities which are qualitatively comparable
to the exact results. We note, however, that if the model Hamiltonian is not supplemented with the confining potential,
the approximate mapping dynamics diverges and no solution is possible. Neither the exact quantum dynamics nor the full QCLE dynamics
suffers from this problem. This indicates that if the mapping dynamics is not confined to the physical space, the instabilities
can probe unphysical regions of models with high probability and spoil the results.

\section{Summary and Comments}\label{sec:conc}
This investigation of the representation of the quantum-classical Liouville equation in the mapping basis led to
several results. From considerations of how the equations of motion and expectation values involve projectors onto
the mapping states corresponding to the physical space, it was demonstrated that the QCL operator commutes with the
projection operator so that the dynamics is confined to the physical space. Further, it was shown that a trajectory-based
solution of this equation entails the simulation of an ensemble of entangled trajectories. The development of suitable algorithms
for the simulation of entangled trajectories is a topic of current research.

The PBME approximation to the QCLE is closely related to the equations of motion in the LSC-IVR approximation to quantum
dynamics~\cite{thoss99,stock05,miller78,miller01,miller07}. It neglects a portion of the back coupling of the quantum subsystem
on its environment and
does admit a solution in terms of an ensemble of independent Newtonian-like trajectories, but the dynamics does not commute
with the projection operator and, thus, the dynamics may take the system outside the physical space. This can lead to unstable
trajectories arising from inverted potentials in the equations of motion. In addition to initially unstable trajectories, dynamical
instabilities can arise in the course of the evolution. As in other studies~\cite{stock05,bonella01,bonella05}, these instabilities
are partially removed by a judicious
choice of mapping Hamiltonian. In this circumstance the PBME equation yields qualitatively, or sometimes quantitatively, accurate results
at small computational cost.

\begin{acknowledgments}
This work was supported in part by a grant from the Natural Sciences and Engineering Council of Canada.
\end{acknowledgments}

\section*{Appendix A: Equivalence of Wigner transforms of products of mapping operators}

In this Appendix we show that Eqs.~(\ref{eq:prod-1stform}) and (\ref{eq:prod-2ndform}) are equivalent.
Denoting the expression in Eq.~(\ref{eq:prod-1stform}) for the matrix product
$A_{W}^{\lambda \nu}(X)B_{W}^{\nu \lambda'}(X)$ by ${\mathcal I}$ and inserting the definition of
$g_{\lambda \lambda'}(x)$ in Eq.~(\ref{eq:g-def}) for the two $g$ factors we obtain
\begin{eqnarray}
&&{\mathcal I}=  \frac{1}{(2 \pi \hbar)^{2N}} \int dx dx' \; A_m(x,X) \int dz dz'\; e^{i(p \cdot z +p' \cdot z')/\hbar}
 \nonumber \\
&&\; \langle r-\frac{z}{2}| r'+\frac{z'}{2}\rangle
 \langle r'-\frac{z'}{2}|m_{\lambda'} \rangle \langle m_{\lambda} | r+\frac{z}{2}\rangle B_m(x',X),
\end{eqnarray}
where we have used completeness on the set of mapping states. Letting $r_c=(r+r')/2$ and $r_r=r-r'$, with a
similar change of variables for the $z$ variables, and using the relation
$\langle r-\frac{z}{2}| r'+\frac{z'}{2}\rangle =\delta(r_r-z_c)$ we find
\begin{eqnarray}
&&{\mathcal I}=   \frac{1}{(2 \pi \hbar)^{2N}} \int dr_c \; \int dp dp' \; \int dz_c dz_r \; 
  A_m(r_c+\frac{z_c}{2},p)
e^{i(p+p') \cdot z_c/\hbar} e^{i(p-p') \cdot z_r)/2\hbar}
  \\
&&\;  \langle r_c+\frac{z_r}{4}-z_c|m_{\lambda'} \rangle \langle m_{\lambda} | r_c+\frac{z_r}{2}+z_c \rangle
B_m(r_c-\frac{z_c}{2},p'). \nonumber
\end{eqnarray}
We have not indicated the dependence on $X$ in this equation. Using the definition of $g_{\lambda \lambda'}$ we can write
\begin{eqnarray}
&&\langle r_c+\frac{z_r}{4}-z_c|m_{\lambda'} \rangle \langle m_{\lambda} | r_c+\frac{z_r}{2}+z_c \rangle = 
 \frac{1}{(2 \pi \hbar)^N} \int d \bar{p}_c \; e^{-i2 \bar{p}_c \cdot z_c/\hbar}
g_{\lambda \lambda'}(r_c+\frac{z_r}{4},\bar{p}_c). \nonumber
\end{eqnarray}
Furthermore,
\begin{equation}
g_{\lambda \lambda'}(r_c+\frac{z_r}{4},\bar{p}_c)=e^{(z_r/4)\cdot \nabla_{r_c}}
g_{\lambda \lambda'}(r_c,\bar{p}_c).
\end{equation}
Inserting these expressions into ${\mathcal I}$, integrating by parts to move the translation operator to the other
functions in the integrand and returning to the $z$ and $z'$ functions, we find
\begin{eqnarray}
&&{\mathcal I}=   \frac{1}{(2 \pi \hbar)^{3N}} \int dr_c d \bar{p}_c \; g_{\lambda \lambda'}(r_c,\bar{p}_c)
\int dp dp'  
dz dz'   e^{i(p- \bar{p}_c) \cdot z/\hbar} e^{i(p'-\bar{p}_c) \cdot z')/2\hbar}
 \nonumber \\ && \quad \times
 A_m(r_c+\frac{z'}{2},p) B_m(r_c-\frac{z}{2},p').
\end{eqnarray}
Next we make use of the Fourier transforms of $A_m$ and $B_m$,
\begin{eqnarray}
&&A_m(r_c+\frac{z'}{2},p)= \int d\sigma d\tau \; e^{i(\sigma \cdot (r_c +z'/2)+\tau \cdot p)/\hbar}
\alpha_m(\sigma, \tau)  \nonumber \\
&&B_m(r_c+\frac{z'}{2},p')= \int d\sigma' d\tau' \; e^{i(\sigma' \cdot (r_c -z/2)+\tau' \cdot p')/\hbar}
 \beta_m(\sigma', \tau')
\end{eqnarray}
Inserting these expressions into the previous form of ${\mathcal I}$, performing the integrals over $z$ and $z'$
to obtain delta functions and finally performing the integrals over $p$ and $p'$, we obtain
\begin{eqnarray}
&&{\mathcal I}=   \int dr_c d \bar{p}_c \; g_{\lambda \lambda'}(r_c,\bar{p}_c) 
\Big[ \frac{1}{(2 \pi \hbar)^{N}}  \int d\sigma d\tau d\sigma' d\tau' \; e^{i(\sigma \cdot r_c+\tau \cdot \bar{p}_c)/\hbar}
\alpha_m(\sigma, \tau) \nonumber \\
&&e^{i(\tau\cdot \sigma'-\tau' \cdot \sigma)/2\hbar}
\beta_m(\sigma', \tau') e^{i(\sigma' \cdot r_c+\tau' \cdot\bar{p}_c)/\hbar}\Big]. \nonumber
\end{eqnarray}
As shown in Ref.~[\onlinecite{imre67}], the quantity in square brackets is $(\hat{A}_m \hat{B}_m)_W$, which establishes the equality between the expressions.

\section*{Appendix B: Evolution operators and projections onto the physical space}
In this Appendix we establish the equality given in Eq.~(\ref{eq:LPcommute}) that shows
$i{\mathcal L}_m$ commutes with the projection operator ${\mathcal P}$. Inserting the definitions of $B_m({\mathcal X})$, $B_m^{{\mathcal P}}({\mathcal X})$, $\rho_m({\mathcal X})$ and
$\rho_m^{{\mathcal P}}({\mathcal X})$ given in Eqs.~(\ref{Bm_sub_wigner}), (\ref{Bm_sub_wigner-proj}), (\ref{eq:unproj-den}) and (\ref{eq:proj-den}), the equation takes the form,
\begin{eqnarray} \label{eq:LPcommuteA}
&&\int dX\; B_W^{\mu \mu'}(X) \Big[\int dx \; c_{\mu \mu'}(x) i{\mathcal L}_m g_{\nu' \nu}(x) \Big]
\rho_W^{\nu \nu'}(X,t) \nonumber \\
&&=\int dX\; B_W^{\mu \mu'}(X) \Big[\int dx \; g_{\mu' \mu}(x) i{\mathcal L}_m c_{\nu \nu'}(x) \Big]\rho_W^{\nu \nu'}(X,t) 
\end{eqnarray}
In Ref.~[\onlinecite{nassimi10}] we showed that
\begin{equation}
\int dx \; g_{\mu' \mu}(x) i{\mathcal L}_m c_{\nu \nu'}(x)= i{\mathcal L}_{\mu' \mu, \nu \nu'},
\end{equation}
so that the right side of Eq.~(\ref{eq:LPcommuteA}) takes the form
\begin{equation}
\int dX\; B_W^{\mu \mu'}(X) i{\mathcal L}_{\mu' \mu, \nu \nu'}  \rho_W^{\nu \nu'}(X,t).
\end{equation}
After an integration by parts with respect to the mapping phase space coordinates, the left side of
Eq.~(\ref{eq:LPcommuteA}) can be written as
\begin{equation}
\int dX\; B_W^{\mu \mu'}(X) i{\mathcal L}^*_{\nu' \nu, \mu \mu'}  \rho_W^{\nu \nu'}(X,t).
\end{equation}
Since $i{\mathcal L}^*_{\nu' \nu, \mu \mu'}=i{\mathcal L}_{\mu' \mu, \nu \nu'}$ (see Eq.~(\ref{eq:QCLop-sub})),
this establishes the identity.

Following a similar strategy we can show that the Poisson bracket mapping operator $i{\mathcal L}_m^{PB}$ does
not commute with  ${\mathcal P}$. To do this we show that
\begin{eqnarray}\label{eq:LPB-P-noncommute}
&&\int d{\mathcal X} \; B_m({\mathcal X}) i{\mathcal L}_m^{PB} \rho_{m}^{\mathcal P}({\mathcal X},t) 
 \ne \int d{\mathcal X} \; B_m^{\mathcal P}({\mathcal X}) i{\mathcal L}_m^{PB}
\rho_{m}({\mathcal X},t). \nonumber
\end{eqnarray}
Since $i{\mathcal L}_m=i{\mathcal L}_m^{PB}+ i{\mathcal L}_m^{\prime}$, it suffices to show that
\begin{eqnarray}\label{eq:Lprime-P-noncommute}
&&\int d{\mathcal X} \; B_m({\mathcal X}) i{\mathcal L}_m^{\prime} \rho_{m}^{\mathcal P}({\mathcal X},t) 
 \ne \int d{\mathcal X} \; B_m^{\mathcal P}({\mathcal X}) i{\mathcal L}_m^{\prime}
\rho_{m}({\mathcal X},t), \nonumber
\end{eqnarray}
and we are again led to consider integrals like those in Eq.~(\ref{eq:LPcommuteA}) except that $i{\mathcal L}_m$ is
replaced by $i{\mathcal L}_m^\prime$. In Ref.~[\onlinecite{nassimi10}], Eq.~(29), we established
\begin{eqnarray}
&&\int dx \; g_{\mu \mu'}(x) i{\mathcal L}_m^{\prime} c_{\nu \nu'}(x)\rho_W^{\nu \nu'}(X,t)
= \frac{1}{4} \delta_{\mu \mu'} {\rm Tr}\Big(\frac{\partial \overline{h}}{\partial R}
\cdot \frac{\partial \rho_W}{\partial P} \Big).
\end{eqnarray}
Evaluation of the corresponding integral using integration by parts gives
\begin{eqnarray}
&&\int dx \; c_{\mu \mu'}(x) i{\mathcal L}_m^{\prime} g_{\nu' \nu}(x)\rho_W^{\nu \nu'}(X,t)
= \frac{1}{4} \delta_{\nu \nu'} \Big(\frac{\partial \overline{h}^{\mu \mu'}}{\partial R}
\cdot \frac{\partial \rho_W^{\nu \nu'}}{\partial P} \Big).
\end{eqnarray}
Thus, the evaluation of Eq.~(\ref{eq:Lprime-P-noncommute}) yields
\begin{eqnarray}
&&\frac{1}{4} \int dX \; {\rm Tr}\Big(B_W \frac{\partial \overline{h}}{\partial R}  \Big) \cdot
\frac{\partial ({\rm Tr}\rho_W)}{\partial P} 
\ne\frac{1}{4} \int dX \; ({\rm Tr}B_W) {\rm Tr}\Big(\frac{\partial \overline{h}}{\partial R}
 \cdot \frac{\partial \rho_W}{\partial P} \Big), 
\end{eqnarray}
establishing the fact that $i{\mathcal L}_m^{PB}$ does not commute with the projection operator ${\mathcal P}$.

\section*{Appendix C: Integration scheme}\label{sec:alg}

We present an integration scheme to solve the system of equations~(\ref{Lm-gen}). This scheme is based on an operator-splitting method, which is motivated by the separation of time-scales between the electronic and nuclear motions in the problem. This method is time-reversible, symplectic, and includes an analytic solution for the quantum subsystem degrees of freedom.

The formal solution to the PBME~(\ref{qclm}) for a dynamical variable $B_m({\mathcal X},t)$ is,
\begin{equation}
 B_m({\mathcal X},t) = e^{i\mathcal{L}_m^{PB} t}B_m({\mathcal X},0),
\end{equation}
and writing a short time decomposition of the propagator we have,
$e^{i\mathcal{L}_m^{PB} t} = \prod_{k=1}^K e^{i\mathcal{L}_m^{PB} \Delta t} $.
The total Hamiltonian in Eq.~(\ref{eq:map-H-trace}) can be written as a sum of two parts, $H_m = H_1+H_2$,
\begin{eqnarray}
\label{eq:H1H2}
H_1 = \frac{P^2}{2M},  \qquad
 H_2 = V_0(R)+\frac{1}{2\hbar} \bar{h}^{\lambda \lambda'}(R) (r_{\lambda} r_{\lambda'}+p_{\lambda} p_{\lambda'}).
\end{eqnarray}
The first part in the decomposition of $H_m$ is chosen to be the
kinetic energy of the environment, and the second part contains the remainder of the terms in the mapping Hamiltonian~(\ref{eq:map-H-trace}).
This choice of decomposition is motivated by the desire to enhance the stability of the approximate integration scheme
and to minimize the difference between the  true Hamiltonian $H$, which
is conserved by the exact dynamics, and the pseudo-Hamiltonian $H_{\text{pseudo}}$, which is exactly conserved by
the approximate dynamics dictated by the integration scheme.

Partitioning the Hamiltonian in this way generates new Liouville operators,
$i \mathcal{L}_j = -\big\{ H_j, \cdot \big\}_{{\mathcal X}}$,
such that
$i \mathcal{L}_0 = i( \mathcal{L}_1+  \mathcal{L}_2)$.
We then express each of the short-time propagators using the symmetric Trotter decomposition,
\begin{equation}
\label{eq:short-Trotter}
e^{i (\mathcal{L}_1+\mathcal{L}_2) \Delta t} =  e^{i \mathcal{L}_1(\frac{\Delta t}{2})}
e^{i \mathcal{L}_2 \Delta t} e^{i \mathcal{L}_1(\frac{\Delta t}{2})} + \mathcal{O}(\Delta t^3).
\end{equation}

The decomposition is most useful if the action of the individual propagators $e^{i \mathcal{L}_i \Delta t}$ on
the phase points of the system can be evaluated exactly.  When this is the case, the exact dynamics of the
integration scheme is governed by the pseudo-Hamiltonian
\begin{eqnarray*}
H_{\text{pseudo}} &=& H + \frac{\Delta t^2}{12} \left( \frac{P}{M} \cdot \frac{\partial^2 H_2}{\partial R \partial R} \cdot
\frac{P}{M} - \frac{1}{2M} \frac{\partial H_2}{\partial R} \cdot \frac{\partial H_2}{\partial R} \right) + O(\Delta t^4).
\end{eqnarray*}
For the decomposition in Eq.~(\ref{eq:H1H2}), the difference between the true Hamiltonian $H$ and the pseudo-Hamiltonian
is of the form of the standard Verlet scheme, and depends only on the smoothness of the effective potential $H_2 (R)$
for the environment variables $R$ and $P$ and not on the smoothness of the phase space variables representing
the quantum subsystem.  This form of the splitting is particularly helpful when the system passes through regions of phase space where there
are rapid changes in the populations of the diabatic quantum states.  For trajectories passing through such regions, which are
common in mixed quantum/classical systems, other decompositions of the Hamiltonian
result in unstable integrators unless very small timesteps $\Delta t$ are chosen.

The evolution under $i \mathcal{L}_1$ gives rise to a system propagator on the environmental coordinates alone,
\begin{equation}
\label{eqLL1-prop}
e^{i\mathcal{L}_1 \Delta t} \left(\begin{array}{c} r(t) \\ p(t) \\ R(t) \\ P(t) \end{array}\right) = \left(\begin{array}{c} r(t) \\ p(t) \\ R(t)+\frac{P(t)}{M}\Delta t \\ P(t) \end{array}\right).
\end{equation}
Evolution under $i \mathcal{L}_2$ looks somewhat more complicated; however, it may evaluated analytically as $R(t)$ is stationary under this portion of the dynamics. The equations of motion are as follows:
\begin{eqnarray} \label{L2}
&& \frac{d r_{\lambda}}{dt}= \frac{ \bar{h}^{\lambda, \lambda'}(R)}{ \hbar}  p_{\lambda'}, \;
\frac{d p_{\lambda}}{dt}= -\frac{\bar{h}^{\lambda, \lambda'}(R)}{ \hbar}  r_{\lambda'}, \;
\frac{d R}{dt} =0 \nonumber \\
&&\frac{d P}{dt}= -\frac{ \partial V_0 (R)}{\partial R} - \frac{1}{2 \hbar}  \frac{ \partial \bar{h}^{\lambda, \lambda'}(R)}{\partial R} (r_{\lambda} r_{\lambda'}+p_{\lambda} p_{\lambda'}).
\end{eqnarray}

Consider the spectral decomposition of the mapping Hamiltonian, \begin{eqnarray}
\bar{h}^{\lambda \lambda'} (R) = C_{\lambda \mu} (R) E_{\mu}(R) C^{-1}_{\mu \lambda'}(R),
\end{eqnarray}
where $E_{\mu}(R)$ are the eigenvalues (adiabatic energies) of $\bar{h}$. The columns of the matrix ${C}$ correspond to the eigenvectors of $\bar{h}$. To simplify the evolution equations for the mapping variables, we use the spectral decomposition of $\bar{h}$ to perform the following transformation,
\begin{eqnarray}
\widetilde{r}_{\lambda} =  C^{-1}_{\lambda \lambda'} r_{\lambda'}, \quad
\widetilde{p}_{\lambda} =  C^{-1}_{\lambda \lambda'} p_{\lambda'}.
\end{eqnarray}

The two coupled equations for $r_{\lambda}$ and $p_{\lambda}$ from Eq.~(\ref{L2}), in the tilde variables, become
\begin{eqnarray}
\label{eq:coupled}
\frac{d \widetilde{r}_{\lambda}}{d t} = \frac{ E_{\lambda} (R) }{\hbar} \widetilde{p}_{\lambda},  \quad
\frac{d \widetilde{p}_{\lambda}}{d t} = \frac{-E_{\lambda} (R)}{\hbar} \widetilde{r}_{\lambda}.
\end{eqnarray}
The above system may be expressed as the matrix equation,
$\frac{d u}{dt} ={\mathcal M} u$,
where the transpose $u^T$ of the vector $u$ for an arbitrary quantum subsystem is written as,
$u^T=(\widetilde{r}_1,\widetilde{p}_1, \cdots, \widetilde{r}_N, \widetilde{p}_N)$.
The matrix ${\bf M}$ has the simple block diagonal form,
\begin{equation}
{\mathcal M} = \frac{1}{\hbar}\bigoplus_{\lambda}
\left(
\begin{array}{cc}
0  &  E_{\lambda}  \\
 -E_{\lambda} &   0
\end{array}
\right) = i \bigoplus_{\lambda} \omega_{\lambda} \sigma_y,
\end{equation}
where  $\omega_{\lambda} (R) = E_{\lambda} (R) / \hbar$, $\bigoplus$ is the matrix direct sum, and $\sigma_y$ belongs to the set of $2\times2$ Pauli matrices.

The general solution to Eq.~(\ref{eq:coupled}) is
$u(t+\Delta t) = e^{ {\mathcal M} \Delta t} u(t)$,
where, in this particular case, the matrix exponential has the form
\begin{equation}
e^{ {\mathcal M} \Delta t}  =  \bigoplus_{\lambda} ( \cos(\omega_{\lambda} \Delta t) 1 + i \sin(\omega_{\lambda} \Delta t)\sigma_y ).
\end{equation}

The time evolved tilde variables are thus obtained,
\begin{eqnarray}
\nonumber  \tilde{r}_{\lambda}(t+\Delta t) &=& \cos(\omega_{\lambda}\Delta t)\tilde{r}_{\lambda}(t) + \sin(\omega_{\lambda} \Delta t)\tilde{p}_{\lambda}(t), \\
  \tilde{p}_{\lambda}(t+\Delta t) &=& \cos(\omega_{\lambda} \Delta t)\tilde{p}_{\lambda}(t) - \sin(\omega_{\lambda} \Delta t)\tilde{r}_{\lambda}(t).
\end{eqnarray}
These results can then be back-transformed to the original (untilded) variables,
and used to solve for the time-evolved momenta from equation~(\ref{L2}).
The explicit form for $P(t+\Delta t)$ is
\begin{eqnarray}
\nonumber P(t+\Delta t) &=&P(t) - \frac{\partial V_0(R)}{\partial R}\Delta t  - \frac{\Delta t}{2\hbar} \frac{\partial E_{\lambda}(R)}{\partial R} (\tilde{r}_{\lambda}(t)^2+\tilde{p}_{\lambda}(t)^2-1).
\end{eqnarray}
Hence, the evolution under $\mathcal{L}_2$ is given by,
\begin{equation}
e^{i\mathcal{L}_2 \Delta t} \left(\begin{array}{c} r(t) \\ p(t) \\ R(t) \\ P(t) \end{array}\right) = \left(\begin{array}{c} r(t+\Delta t) \\ p(t+\Delta t) \\ R(t) \\ P(t+\Delta t) \end{array}\right).
\end{equation}

\newpage

\begin{thebibliography}{58}
\bibitem[{foo({\natexlab{c}})}]{footnote:PBsims}
\bibinfo{note}{Although the structure of the evolution equations for the phase
  space variables are the same, differences can arise from the specific form of
  the Hamiltonian, e.g., whether the trace is or is not removed, and the
  sampling used in the evaluation of the expectation value.}
\bibitem[{foo({\natexlab{d}})}]{footnote:two-peaks}
\bibinfo{note}{The two peaks are not fully resolved for $P_0=20$ within the
  statistics used in the Trotter-based simulation for this parameter value.}
\bibitem[{foo({\natexlab{e}})}]{footnote-notation}
\bibinfo{note}{We adopt the standard notation for this model using $X$ and $Y$,
  and corresponding momenta $P_X$ and $P_Y$, for the tuning and copuling
  coordinates. Thus, in the general notation of the text $R_s=(X,Y)$ and
  $P_s=(P_X,P_Y)$. The full bath phase space coordinate $X=(R,P)$ should not be
  confused with the tuning coordinate of the FLV model.}
\bibitem[{foo({\natexlab{f}})}]{footnote:RM}
\bibinfo{note}{The full two-state quantum system was simulated with and without
  $V_a$. Solution of the dynamics requires the evolution of the wave function
  $\Psi_\lambda(X,Y)$, since $X$ and $Y$ are now quantum mechanical operators.
  The numerical procedure was based on a discretization of ($X$, $Y$) space and
  a Trotter decomposition of the propagator into a free part and a potential
  part. The free part is solved using discrete fast Fourier transforms for each
  of the two levels, while the potential part involves a per-lattice-site
  $2\times2$ unitary transform between the levels. The initial wave function
  was taken to be a Gaussian wave packet for diabatic level 1 centered around
  $(x_0,y_0)=(5.0494 a_0, 16 a_0)$, with square widths $0.5 \; a_0^2$ in both
  directions. Diabatic level 2 was initially unoccupied. The wave packet was
  given a non-zero $Y$-momentum of $-22$ by adding a phase factor, so that the
  wave packet was directed towards the reaction region.}
\end{thebibliography}

\end{document}